\documentstyle[12pt,epsfig]{ioplppt_my}

\newcommand {\threejot}[6]{\pmatrix{ #1\!\!&#2\!\!&#3\!\cr #4&#5&#6  \cr}}
\newcommand {\sixjot}[6]
{\left\{
\begin{array}{ccc}
\!\! #1\!\! & #2\!\! & #3\! \\
\!\! #4\!\! & #5\!\! & #6\!
\end{array}
\right\} }

\begin{document}
\jl{2}

\title{Formalism for Multiphoton Plasmon Excitation
in Jellium Clusters.}

\author{Jean-Patrick Connerade \dag\ftnote{4}{E-mail:
j.connerade@ic.ac.uk} and
Andrey V Solov'yov\ddag\ftnote{3}{E-mail:
solovyov@rpro.ioffe.rssi.ru}
}

\address{\dag  The Blackett Laboratory, Imperial
College, London SW7 2BZ, UK  }
\address{\ddag A.F.Ioffe Physical-Technical Institute of the Academy
of Sciences of Russia, Polytechnicheskaya 26, St. Petersburg 194021,
Russia}

\begin{abstract}
We present a new formalism
for the description of multiphoton plasmon excitation
processes in jellium clusters.
By using our method, we demonstrate that, in addition
to dipole plasmon excitations,
the multipole plasmons  (quadrupole, octupole, etc)
can be excited in a cluster by multiphoton absorption processes,
which results in a significant difference between plasmon resonance
profiles in the cross sections for multiphoton
as compared to  single-photon absorption.
We calculate the cross sections for multiphoton absorption
and analyse the balance between the surface
and volume plasmon contributions to multipole plasmons.
\end{abstract}

\section{Introduction}

In the present paper we demonstrate that, in addition to
dipole plasmon excitations
multipole plasmons  (quadrupole, octupole,  etc)
contribute to the multiphoton excitation process,
which results in a significant difference of plasmon resonance
profiles between the cross sections for
multiphoton and
single-photon absorption.
We have developed a formalism from which the cross sections
for multiphoton excitation can be worked out.
The balance between multipole surface
and volume plasmon contributions is analyzed.
Our results are obtained
within a theoretical model for the multiphoton excitation of
a jellium cluster.  This model is applicable  to metal clusters,
to fullerenes and to any type of cluster in which a strong
delocalization of valence atomic orbitals takes place.
The theoretical formalism we have developed is not confined in
its application to photons. It can also be used to describe
any kind of higher order plasmon excitation processes, for example
those which arise by multiple scattering of electrons within the cluster.

Recently, a number of papers have discussed metallic clusters
\cite{koller,ullrich} and fullerenes \cite{hunsche}
in strong laser fields. The theoretical approach usually followed is to solve a
time-dependent local density equation numerically (LDA or local-density
approximation), and the regime most commonly
studied involves intense, short laser pulses, for which the turn-on and turn-off
properties significantly affect excitation.

We report here on a different problem: our initial interest lies in lower laser
powers, for which multiphoton excitation just begins to intrude, and we are
interested in developing a formalism to describe the interaction between
collective modes and a laser field in the multiphoton regime. For this purpose,
a semiclassical model, in which the collective flow of charge is driven by a
periodic field, is established, and we relate it to the multiphoton
absorption cross section of the cluster, which takes account of quantum
mechanics. Of course, one can in principle extend this treatment to consider the
turn-on and turn-off of laser pulses, to treat the interaction numerically for
various power levels and initial charge distributions. What we wish to point
out, however, is a novel feature, which arises even for an infinite wavetrain
interacting with a cluster (the simplest and most fundamental problem): multiple
plasmon excitations are driven by multiphoton excitations. In the present
paper, we explain by what mechanism this arises.

Our approach is based on different principles from the LDA. Instead of using the
Kohn-Sham formalism, we propose a hydrodynamic approach. In the LDA, all
quantities are made to depend solely on charge density, and currents are
subsequently made to appear by solving a time-dependent equation. In our
approach, we start out from the continuity equation and the Euler equation,
from which both the current flow and the density are obtained
in a completely consistent
way. Our theory is local, and does not include exchange. In principle, it would
be possible to extend it by building in exchange and correlations in a similar
way to the LDA. One of the benefits of our approach is that all the momentum
transfer terms are included in the formulation, which leads to the presence of
both volume and surface plasmon terms. If one wishes to simplify the theory, it
is possible towards the end of the calculation to assume zero momentum transfer,
in which case the volume plasmon terms disappear from the problem.

Surface plasmon excitations are well known in atomic
cluster physics. The dipole surface plasmons
are responsible for the formation of
giant resonances in  photoabsorption spectra of metal clusters
(see e.g.
\cite{deHeer93,Brack93,Brechignac94,Haberland94,Alasia94,Madjet95,Kreibig95,C20_C60,MetCl99}).
They also play an important role in
inelastic collisions of charged particles with metal clusters
\cite{LesHouches00,Korol97,Ekardt86,Ekardt87,Gerchikov97a,Gerchikov97b,Gerchikov98}.
The role of surface
plasmon excitations in inelastic electron-cluster scattering
was thoroughly studied
in \cite{LesHouches00,Gerchikov97a,Gerchikov97b,Gerchikov98}, and
it was demonstrated
that  collective
excitations make a significant contribution to the electron energy loss
spectrum (EELS) in the region of the surface plasmon resonance.
With increasing  scattering angle, plasmon excitations
of higher angular momenta become more and more prominent.

Plasmons
are characteristic of delocalised electrons, and therefore the
jellium model provides the most appropriate starting point
for a discussion of plasmon excitation in the multiphoton regime.
This regime is particularly appropriate for the study of atomic
clusters, which are rather fragile objects, and readily explode
under very strong irradiation \cite{ditmire}

The inelastic scattering of fast electrons on metal clusters
in the range of
transferred energies above the ionization threshold
was considered
in \cite{ioni_cl}.
It was demonstrated that, in
this energy range,  volume plasmons  dominate the  contribution to
the differential cross section, resulting in a resonance behaviour.
The volume plasmon resonances excited in the cluster
during a collision decay via the
ionization process. The
resonance frequency and the autoionization width of the volume plasmon
excitations have both been determined  in \cite{ioni_cl}.

The role of the polarization interaction and
plasmon excitations in the process of electron attachment
to metal clusters has also been examined both
theoretically \cite{at_let,at_pap}
and experimentally \cite{at_exp}. It was demonstrated that
plasmon excitations induce a resonance enhancement
of the electron attachment cross section.

Our paper is organized as follows. In section \ref{m_phot},
we derive quantum mechanical expressions for
the cross sections
of multipole (quadrupole, octupole etC) plasmon excitations
taking place in the multiphoton absorption regime
and estimate the cross sections on the basis of the plasmon
resonance approximation.
We present and discuss
the multiphoton absorption profiles and
demonstrate
a significant change of the multiphoton
absorption profiles in the cross sections for
multiphoton as compared to
single-photon absorption.
In section \ref{hydrodyn}, we establish certain connections
of the cross sections
with the variation of electron density in the cluster
due to the external field and
present the formalism
for the calculation of the electron density variation in
the cluster due to the external field, based on the use of
the hydrodynamic Euler equation and on the equation
of continuity. We apply the general formalism to the
description  of
fast electron-cluster scattering and multiphoton
absorption. In section \ref{q_mom}, we
calculate the multipole moments of the system induced by
the external field on the basis of the formalism outlined
in section \ref{hydrodyn}.
We analyse the plasmon resonance structure
of the induced multipole moments and conclude
that it is analogous to the one arising in
the multiphoton absorption cross sections calculated
in section \ref{m_phot}.
In section \ref{conclu}, we draw conclusions from this work.
In \ref{coll_matr_el}, we show how matrix elements
for collective transitions can be calculated on the
basis of the sum rules.
In \ref{integrals}, we present details of calculations
of the angular integrals arising in the
formalism outlined here.

\section{Plasmon resonance approximation for
multiphoton absorption cross sections}
\label{m_phot}

First, we consider the cross section for
multiphoton absorption in jellium clusters and
demonstrate that multipole plasmon
excitations are essential to this process.
The discussion in this section is based on the
plasmon resonance approximation, which is introduced
below.

\subsection{Single photon absorption}

Let us start by  considering the
simplest example and calculate
the  cross section for single-photon absorption in
the plasmon resonance approximation.

The single-photon absorption cross
section in the dipole approximation reads as
\begin{equation}
\sigma_1= \frac{4\pi^2 e^2  }{ c} \omega \sum_n  |z_{on}|^2
\delta(\omega_{no} -\hbar\omega)
\label{sigma_1}
\end{equation}

Here, $e$ is the charge of electron, $c$ is the velocity of light,
$\hbar$ is  Planck's constant,  $\omega_{no}=\varepsilon_n -
\varepsilon_0$ is the electron excitation energy,  $\omega$ is the
photon frequency and $e z_{on}$ is the matrix element of the $z$-
component of the cluster dipole moment.
The summation over $n$ includes all
final  states  of the excited electron, which belong
to both the discrete and the continuous spectra.

In the jellium picture, which works reasonably well for
metal clusters and to some extent for fullerenes,
the main contribution to the cross section
(\ref{sigma_1}) arises  from a small group of excited states or
sometimes even from a single transition of frequency
close to the classical Mie resonance frequency -- also
known as the frequency of the plasmon resonance.
For a spherical
metal cluster, this frequency is given by
(see e.g. \cite{Kreibig95,Gerchikov97a} and section 4 of this paper)
\begin{equation}
\omega_l^2 = \frac{4 \pi N e^2}{mV} \cdot \frac{l}{(2l+1)}.
\label{omega_pl}
\end{equation}

Here $V= 4\pi R^3/3 $ is the cluster volume, where $R=r_o N^{1/3}$ is the
cluster radius, $r_o$ is the Wigner-Seitz radius; $N$ is the number
of delocalized electrons in a cluster, $l$ is the angular momentum
of the plasmon mode, $m$ is the electron mass. Note that, by using a single
photon of energy
$1-4 eV$, one can, in practice, excite only $l=1$ dipole plasmon oscillations
in a metallic cluster.

For nearly-spherical fullerenes $C_{20}$ or $C_{60}$,
the plasmon resonance frequency is equal to
(see \cite{C20_C60,Gerchikov97a})
\begin{equation}
\omega_l^2=\frac{l(l+1)N}{(2l+1)R^3}
\label{freq_ful}
\end{equation}
\noindent
where $N$ is the total number of delocalised electrons (4 electrons per atom
times the number of carbon atoms in the fullerene molecule) and $R$ is the
the radius of the fullerene.

The plasmon resonance approximation is based on the
fact that excitations in the vicinity of a plasmon resonance exhaust
the sum rule almost completely
(see \cite{deHeer93,Brack93,Brechignac94,Haberland94,Kreibig95,C20_C60}),
which means that the summation in the sum rule (see e.g. \cite{LL3})
\begin{equation}
\sum_n \omega_{no} |z_{on}|^2  = \frac{N\hbar^2 }{2m}
\label{sum_rule}
\end{equation}
\noindent
need be performed only over excited states
in the vicinity of
the Mie resonance.

Now, assuming a Lorentzian distribution
of width $\Gamma_1$
for the plasmon resonance states
and
replacing the delta function, $\delta(\omega_{no} -\hbar\omega)$,
in (\ref{sigma_1})
by the profile (see e.g. \cite{LL3})
\begin{equation}
\delta(\omega_{no} -\hbar\omega) \longrightarrow
\frac{\Gamma_1}
{2\pi\hbar
((\omega_1-\omega)^2 + \frac{\Gamma_1^2}{4} )},
\label{Lorentz}
\end{equation}
one recovers
the well-known expression for
the single-photon absorption cross section
(see e.g. \cite{deHeer93,Kreibig95})
\begin{equation}
\sigma_1=
\frac{\pi N e^2  }{m c}
\frac{\Gamma_1} {(\omega_1  -\omega)^2 + \frac{\Gamma_1^2}{4}}
\approx
\frac{4\pi N e^2  }{m c}
\frac{\omega^2 \Gamma_1 } {(\omega_1^2  -\omega^2)^2 + \omega^2\Gamma_1^2}
\label{photo_1}
\end{equation}

The width  $\Gamma_1$ is due to Landau damping.
Its calculation for metal clusters is performed,
for example, in \cite{ioni_cl}.

The cross section (\ref{photo_1}) reproduces correctly
the appearence of the plasmon resonances in single-photon
absorption spectra of metal clusters and fullerenes,
although some details of the experimentally observed profiles
are naturally beyond the plasmon resonance approximation.
The discussion of these details is not the scope of the
present work. This can only be done accurately enough
on the basis of {\it ab initio} many-body theories. Instead,
we analyse the multiphoton absorption cross sections on
the basis of the plasmon resonance approximation and elucidate
the role of multipole plasmon excitations in its formation,
because our interest lie in establishing the physical mechanisms
which underpin multiphoton excitation.

\subsection{Two-photon absorption}

In the dipole approximation, the two-photon absorption cross
section is equal to
\begin{equation}
\sigma_2= \frac{32\pi^3 e^4 \hbar  }{c^2} \omega^2  \sum_n
\left|\sum_m \frac{z_{nm}z_{mo}}{\hbar\omega-\omega_{mo}+\i\delta}\right|^2
\delta(\omega_{no} -2\hbar\omega)
\label{sigma_2}
\end{equation}

We evaluate the cross section (\ref{sigma_2}) in the same way as for
the single-photon case.  The main contribution
to the sum over the intermediate states $m$  arises
from
the virtual dipole plasmon excitations. Therefore,
one derives
\begin{equation}
\sum_n
\left|\sum_m \frac{z_{nm}z_{mo}}{\hbar\omega-\omega_{mo}+\i\delta}\right|^2
\approx
\frac{N }{2 m \hbar \omega_1} \sum_n
\frac{|z_{n1}|^2 } {(\omega  -\omega_1)^2 + \Gamma_1^2/4}
\label{estim1}
\end{equation}

Here, we have also introduced a dipole plasmon resonance width $\Gamma_1$
and used the sum rule (\ref{sum_rule}) for the evaluation of the
matrix elements
for the dipole plasmon excitation $|r_{10}|^2 \approx  N\hbar/2m\omega_1$.
The remaining matrix elements $z_{n1}$ in (\ref{estim1}) describe
dipole transitions
from the dipole plasmon resonance state
to  other excited states.
Matrix elements for these transitions obey
the dipole selection rule. This means that
the angular momentum of the final state can only be equal to either
$l=0$ or $l=2$. According to (\ref{omega_pl}) and (\ref{freq_ful}),
there is no surface plasmon excitation with $l=0$  either in
metal clusters or in fullerenes. Thus, only transitions to
the states with $l=2$ are of interest.

These arguments show that, by using two photons simultaneously,
one can excite the quadrupole plasmon resonance in a metal cluster
or in a fullerene with a frequency given in (\ref{omega_pl}) and
(\ref{freq_ful}) respectively. When calculating the cross section
(\ref{sigma_2}) in the vicinity of the quadrupole plasmon resonance
excitation, i.e. at $2\omega \sim \omega_2$,  it is sufficient to
consider only transitions to the resonance final state,
i.e. to put
$\sum_{n} |z_{n1}|^2 \approx |z_{21}|^2$ (here and below
we use indices 1 and 2 to designate the dipole and quadrupole
plasmon resonance states) and
to replace the delta function, $\delta(\omega_{no} - 2\hbar\omega)$
by a Lorentzian distribution
of width $\Gamma_2$ (see e.g. \cite{LL3})
\begin{equation}
\delta(\omega_{no} -2\hbar\omega) \longrightarrow
\frac{\Gamma_2}
{2\pi\hbar
((\omega_2- 2\omega)^2 + \frac{\Gamma_2^2}{4} )}.
\label{Lorentz2}
\end{equation}

By substituting (\ref{estim1}) and (\ref{Lorentz2}) in (\ref{sigma_2}),
one derives

\begin{eqnarray}
\sigma_2 =
\left(\frac{4\pi N e^2 }{m c}\right)^2
\frac{m |z_{21}|^2}{2 N  \hbar\omega_1 }\cdot
\frac{\omega^2 } {(\omega- \omega_1)^2 +
\frac{\Gamma_1^2}{4}} \cdot
\frac{\Gamma_2 } {(\omega_2  -2\omega)^2 +
\frac{\Gamma_2^2}{4}}
\label{photo_2a}
\end{eqnarray}

This result demonstrates that the photoabsorption profile in the two-photon
process differs substantially from
the single-photon case. The cross section
(\ref{photo_2a})  has a resonance at
the dipole plasmon frequency and, in addition, also contains the
quadrupole plasmon resonance
at $\omega = \omega_2/2$.

The cross section (\ref{photo_2a}) depends on
$|z_{21}|^2$. The transition matrix element $z_{21}$
describes the electron transition between the dipole
and quadrupole plasmon resonance states. This is
a single electron transition rather than a
collective one. Therefore, calculation of $z_{21}$ on the basis of the
sum rule (\ref{sum_rule}) would lead to a significant overestimate
of the value of this matrix element. Instead, one can use
Heisenberg's uncertainty
principle for the evaluation of $z_{21}$ \cite{Migdal}.
By estimating the radial component of the momentum of a single electron in a
dipole and quadrupole plasmon oscillatory mode as
$p_1 \sim m \omega_{1} \Delta R $ and  $p_2 \sim m  \omega_{2} \Delta R$
respectively,
one derives
\begin{equation}
|z_{21}| \sim min(\Delta z_1, \Delta z_2 ) \sim
\frac{\hbar}{p_{2}} \sim
\sim A \frac{\hbar}{m  \omega_{1} \Delta R}
\label{estim2}
\end{equation}

Here, $A$ is a dimensionless constant, of the order of one,
$\Delta z_1 \sim \hbar/p_1$ and $\Delta z_2 \sim \hbar/p_2$
are the uncertainties relating to an electron in the dipole and quadrupole
plasmon modes respectively, $\Delta R$ is the width of the domain in
the vicinity of the cluster surface within which plasmon excitations
take place.  In \ref{coll_matr_el} we prove the correctness of this
estimate and demonstrate that  matrix element $z_{21}$ is equal to
\begin{equation}
z_{21} = - \frac{8}{3}  \left(\frac{6}{5}\right)^{1/4}
\frac{\hbar}{m  \omega_{1} \Delta R}
\label{z12}
\end{equation}

By substituting (\ref{z12}) in (\ref{photo_2a}),
we obtain the final expression for the two-photon
absorption  cross section:
\begin{eqnarray}
\sigma_2 =
\left(\frac{4\pi N e^2 }{m c}\right)^2
\frac{A^2 \hbar}{2m \omega_1 N \Delta R^2 }\cdot
\frac{\omega^2 }{\omega_1^2}
\cdot
\frac{1} {(\omega- \omega_1)^2 +
\frac{\Gamma_1^2}{4}} \cdot
\frac{\Gamma_2 } {(\omega_2  - 2\omega)^2 +
\frac{\Gamma_2^2}{4}}
\label{sigma_2fin}
\end{eqnarray}
\noindent
where $A= \frac{8}{3} \left(\frac{6}{5}\right)^{1/4}
\approx 2.79$.

\begin{figure}[t]
\vspace*{0.cm}
\begin{center}
\includegraphics[scale=0.6]{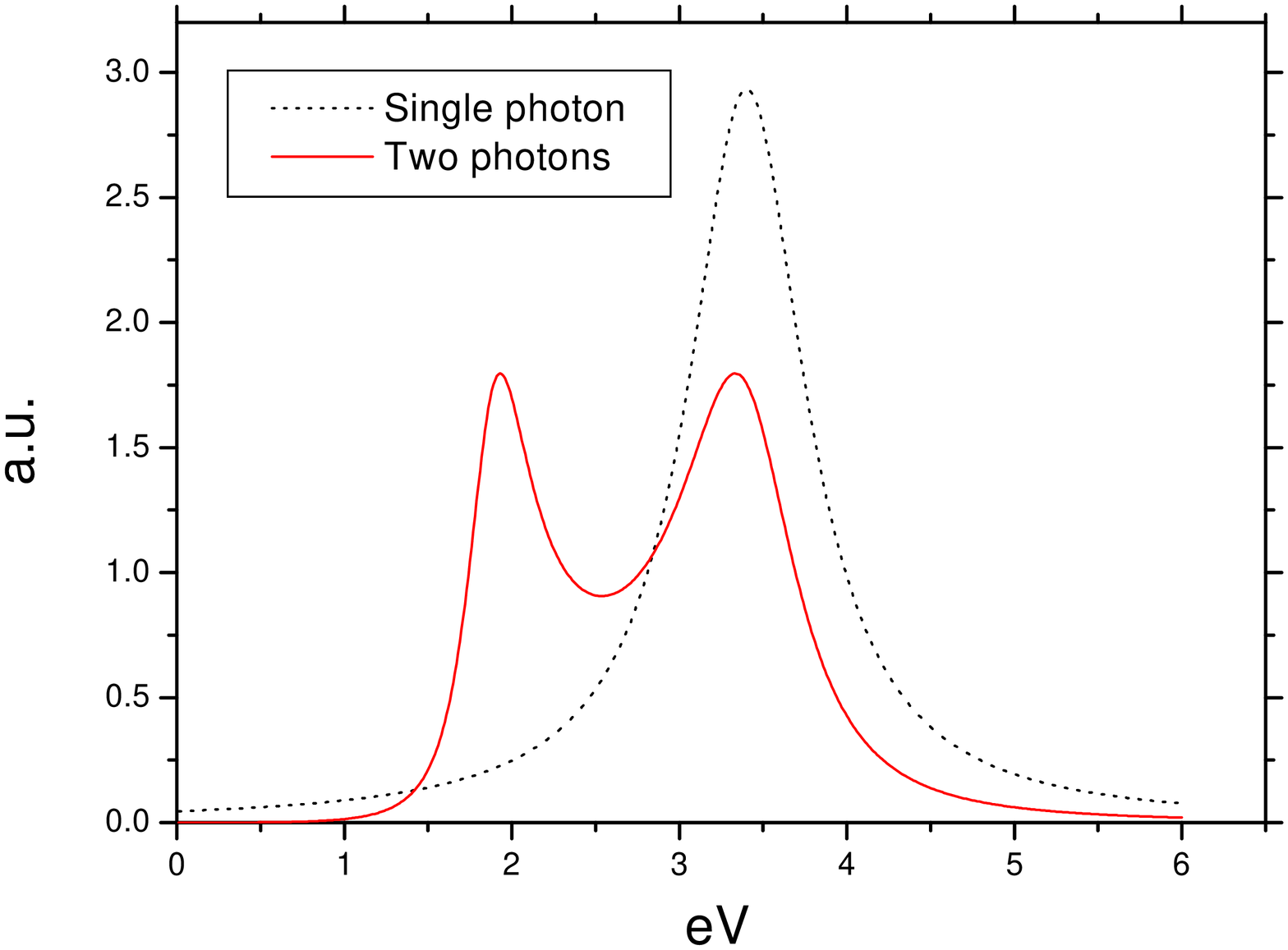}
\end{center}%
\caption{
The profiles of single-photon (dotted line)
and two-photon (solid line) absorption calculated
according to (\protect\ref{photo_1}) and (\protect\ref{sigma_2fin})
and normalised per unit atom. The two-photon absorption profile
is scaled  by a factor 1/100.
The scales are not
identical for the two curves for reasons of definition of the cross sections in
the single- and two-photon cases, but both are given in atomic units.
}
\label{fig1}
\end{figure}

We note that
the cross section (\ref{sigma_2fin}) depends explicitly
on  Planck's constant $\hbar$, while
the cross section (\ref{photo_1}) does not.
The independence of (\ref{photo_1}) from $\hbar$
is connected with
the fact that plasmon oscillations are a purely classical effect,
while the dependence of (\ref{sigma_2fin}) on $\hbar$ arises from
the interaction between dipole and quadrupole plasmon modes
as can be seen from estimate (\ref{estim2}) and the
explicit expression (\ref{z12}).
This indicates that it is meaningful to treat
plasmon excitations  classically, while the coupling of various
plasmon modes in the multiphoton photoabsorption
process must be treated beyond
purely classical theory.

In figure 1 we plot the cross section profiles per unit atom for single-photon
(dashed line) and two-photon (solid line) absorption calculated
according to (\ref{photo_1}) and (\ref{sigma_2fin}).
These profiles do not depend on the number of atoms in the cluster.
Note that the scales are not
the same for the two curves for reasons of definition of the cross sections in
the single- and two-photon cases, but are both given in atomic units. The peak
in the single-photon plot gives the location of the dipole resonance. The other
peak in the two-photon plot is the quadrupole resonance.
This figure demonstrates a significant difference
between the nature of the profiles, arising from the presence of
quadrupole plasmon excitation in the two-photon
case. In this calculation
we have input $r_0=4.0$ and $\Gamma_1=\omega_1/4$, $\Gamma_2=\omega_2/4$,
$\Delta R=r_0$. The choice of these parameters can be different for
different clusters, but it should always lead to qualitatively similar
single- and two-photon absorption profiles. An accurate determination
of the parameters is only possible on the basis of {\it ab initio}
calculations.

\subsection{$n$-photon absorption}

The formalism we have developed can also be used for the evaluation
of the multiphoton absorption cross sections for a larger
number of photons.
In the dipole approximation the n-photon absorption  cross section
has the following form:
\begin{equation}
\sigma_{n}= \frac{(2\pi)^{n+1} n!^2  e^{2n} \hbar^{n-1}  }{c^{n}}
\omega^{n}
\sum_k \left|M_{k}\right|^2 \delta(\omega_{ko} -n\hbar\omega)
\label{sigma_2n}
\end{equation}
\noindent
Here the amplitude $M_n$ is equal to
\begin{eqnarray}
M_{k}&=&
\sum_{m_{n-1}} \sum_{m_{n-2}}.... \sum_{m_1}
\frac{z_{km_{n-1}}}
{
((n-1)\hbar\omega-\omega_{m_{n-1} o}+\i\delta)}\cdot
\nonumber
\\
&\cdot&\frac{
z_{m_{n-1}m_{n-2}}}
{((n-2)\hbar\omega-\omega_{m_{n-2} o}+\i\delta)}
....
\frac{z_{m_{1}0} }
{
(\hbar\omega-\omega_{m_1 o}+\i\delta)
}
\label{M_k}
\end{eqnarray}

The plasmon resonance structure of the multiphoton
absorption cross section (\ref{M_k})
can be analysed in a similar way to
the previous treatments for the single- and two-photon
cases. This analysis immediately leads to the
important conclusion that plasmon resonances with larger
angular momenta (octupole, etc) can be excited
in the multiphoton absorption regime.
Thus, for example,
with three photons, the
octupole plasmon resonance at $\omega = \omega_3/3$
will also be excited.
This analysis, however, leaves undefined the
matrix elements for electronic transitions between
various plasmon modes. Estimates of these can be performed
either on the basis of Heisenberg's Uncertainty Principle
or by a calculation
similar to the one for $Z_{12}$
(see (\ref{z12}) and (\ref{photo_2a})) in the two-photon case, but
their accurate evaluation is not trivial.

Note that the plasmon resonance approximation
allows one to analyse only the plasmon resonance excitations that
are characterised by relatively low angular momenta, because
electron excitations in the cluster with large
angular momenta $l$  have single-particle character.
This follows, for instance,
from the fact that with increasing
$l$ the wave length of the surface plasmon mode, $2\pi R/l$, becomes
smaller than the characteristic wave length of the delocalised electrons
at the Fermi surface, $2\pi \hbar/\sqrt{2m\Delta\varepsilon}$, where
$\Delta\varepsilon$ is the characteristic
electron excitation energy in the cluster.
In other words, excitations with angular momenta comparable with
the characteristic electron angular momenta of the ground state
exhibit
single-particle rather than collective character.
Therefore, when analyzing contributions of the plasmon resonance modes
to the multiphoton absorption cross section,
one should consider only the lowest angular momenta.
For example according to the jellium model,
the maximum angular momentum of the delocalised electrons in the $Na_{40}$
cluster is equal to 4. Therefore, only dipole and quadrupole collective
modes can be expected in this case. With increasing cluster
size the number of essential plasmon modes grows as $R$.

\section{Hydrodynamic description of collective motion of the electron density
in a cluster}
\label{hydrodyn}

The multipole plasmon resonances arising in the multiphoton
absorption cross sections, should also appear in other
physical characteristics of the cluster, which can be probed
in the multiphoton absorption regime.
In the situation where plasmon resonance excitations
are the dominant contribution to
the multiphoton absorption cross section,
it is natural to seek and analyse the plasmon resonance structure
of the variation of  electron density induced by the radiation
field.
The variation
of electron density is a characteristic of the system,
allowing one easily to connect classical and quantum
descriptions of the excitation process, because charge density variation
has the same meaning in
both quantum and  classical mechanics.
A classical description of the electron density variation
in a cluster is appropriate in the situation where
plasmon excitations dominate over the single-particle spectrum, because
plasmon oscillations in clusters are an essentially classical effect.

\subsection{Basic equations}

Since our object of interest is the
excitation of plasmons in metallic clusters,
which have a distinctly classical nature,
we now describe
the collective motion of the electron
density using  Euler's equation and the
equation of continuity.

Euler's equation  couples the acceleration of the electron density
$d{\bf v}/dt$
with the total local electric field ${\bf E}$
acting on the density at the point $({\bf r},t)$.
It has the following form:
\begin{equation}
\frac{d{\bf v}({\bf r},t)}{dt}= \frac{e}{m} {\bf E}({\bf r},t)
\label{Euler1}
\end{equation}
The electric field ${\bf E}$ includes both the external field
acting on the cluster and the polarization contribution
arising from the variation of electron density.
Expressing the total
derivative on the left hand side of (\ref{Euler1}) as
the sum of two contributions, arising from the
change in velocity of the electron density
in time and in space, one obtains:

\begin{equation}
\frac{\partial {\bf v}({\bf r},t)}{\partial t}+
\{{\bf v}({\bf r},t)\cdot {\bf \nabla}\} {\bf v}({\bf r},t)
=
-\frac{e}{m} {\bf \nabla} \varphi ({\bf r},t)
-\frac{e}{m} {\bf \nabla}
\int  d{\bf r^\prime}
\frac{\delta \rho({\bf r^\prime},t)}{|{\bf r}-{\bf r^\prime}|}
\label{Euler2}
\end{equation}

Here $\varphi ({\bf r},t)$ is the potential of the external
field. The second term on the right hand side of (\ref{Euler2})
describes the polarization force  due to the
variation of electron density $\delta \rho({\bf r},t)$.

We assume that the external potential
$\varphi ({\bf r},t)$ is the solution of the wave
equation. Therefore, we can put
\begin{equation}
\varphi ({\bf r},t)= \e^{i\omega t} \varphi ({\bf r}),
\label{varphi}
\end{equation}
where $\varphi ({\bf r},t)$ satisfies the equation
\begin{equation}
\Delta \varphi ({\bf r})= - k^2 \varphi ({\bf r}).
\label{varphi_r}
\end{equation}
Here $k= \omega/c$, $c$ is the velocity of light,
but in principle
one can postulate a more
complex dispersion law.
We need consider only the positive
frequency solution of the wave equation, because the formalism for
the negative frequency solution is analogous to it.

The total electron density in the cluster is equal to
\begin{equation}
\rho({\bf r},t)= \rho_o({\bf r}) + \delta \rho({\bf r},t),
\label{density}
\end{equation}
\noindent
where $\rho_o({\bf r})$ is the electron density distribution
in a free cluster
without an external field and $\delta \rho({\bf r},t)$
is the variation of electron density caused by
the external field and the polarization force acting together.

The motion of  electron density in the cluster obeys the equation
of continuity, which reads:
\begin{equation}
\frac{\partial  \rho({\bf r},t)}{\partial t}+
{\bf \nabla}\cdot \{\rho({\bf r},t) {\bf v}({\bf r},t)\}=0
\label{eq_cont}
\end{equation}

The simultaneous solution of equations (\ref{Euler2}), (\ref{density})
and  (\ref{eq_cont})
with  appropriate initial conditions and the initial distribution
$\rho_o({\bf r})$
allow one to determine the variation of electron density
$\delta \rho({\bf r},t)$
as well as its velocity  ${\bf v}({\bf r},t)$.
We  solve this problem by using
a perturbative approach on the external field
$\varphi ({\bf r},t)$.

\subsection{Perturbation theory}

It is easy to estimate the relative value of the
first and the second terms on the left hand side
of (\ref{Euler2}). We see that  the second term is
negligible, provided the condition
$E \ll m \omega^2 R/e$ is fulfilled.
Substituting here the characteristic values
$\omega \sim 0.1$, $R \sim 10$, one derives $E \ll 0.1$ in
atomic units or  $E \ll 5 \cdot 10^8 B/cm$. Below, we
assume that this condition is fulfilled and neglect
the second term on the left hand side of
(\ref{Euler2}), which means physically that the external
field causes only a small spatial inhomogeneity
in the electron density distribution within the cluster. In this limit,
Euler's equation reduces to a Newtonian equation,
which describes electronic motion in the cluster under
the action of the external field and the polarization force.

We express the solutions of (\ref{Euler2}) and
(\ref{eq_cont}) in the following form:
\begin{equation}
\delta \rho({\bf r},t)= \sum_{n=1}^\infty
\delta \rho_n({\bf r}) \e^{i n \omega  t}
\label{series_rho}
\end{equation}
\begin{equation}
{\bf v}({\bf r},t)= \sum_{n=1}^\infty
{\bf v}_n({\bf r}) \e^{i n\omega  t}
\label{series_v}
\end{equation}

By substituting these expansions into (\ref{Euler2}), (\ref{eq_cont}) and
performing
simple transformations, one derives
\begin{equation}
{\bf v}_n({\bf r})=
\frac{ie}{m n\omega }\delta_{n,1} {\bf \nabla}\varphi({\bf r})+
\frac{ie}{m n\omega }
{\bf \nabla}\int d{\bf r^\prime}
\frac{\delta \rho_n({\bf r^\prime})}{|{\bf r}-{\bf r^\prime}|}
\label{v_n}
\end{equation}

\begin{equation}
i\omega n \delta\rho_n({\bf r}) +
{\bf \nabla} \cdot \Biggl\{\rho_0({\bf r}) {\bf v}_n({\bf r})\Biggr\} +
\sum_{k^\prime=1}^{n-1}
{\bf \nabla} \cdot  \Biggl\{\delta \rho_{k^{\prime}}({\bf r})
{\bf v}_{n-k^{\prime}}({\bf r})\Biggr\} =0
\label{rho_n_1}
\end{equation}

Here $\delta_{n,1}$ is the Kroneker symbol.
One can exclude ${\bf v}_n({\bf r})$ from equation (\ref{rho_n_1}) by
the substitution of (\ref{v_n}) in (\ref{rho_n_1}).
Performing this transformation with the simultaneous use of
(\ref{varphi_r}) and
$\Delta |{\bf r}- {\bf r}^{\prime}|^{-1}=
-4\pi\delta({\bf r}- {\bf r}^{\prime})$,
one derives
the following equation:
\begin{eqnarray}
\left((\omega n)^2 - \frac{4\pi e}{m} \rho_o({\bf r})\right)
\delta \rho_n({\bf r}) +
\frac{e}{m} {\bf \nabla}  \rho_o({\bf r}) \cdot
{\bf \nabla}
\int d{\bf r^\prime}
\frac{\delta \rho_n({\bf r^\prime})}{|{\bf r}-{\bf r^\prime}|}=
\nonumber
\\
=\frac{e}{m}\delta_{n,1}
\left(\rho_o({\bf r})\varphi({\bf r}) k^2 -
{\bf \nabla} \varphi({\bf r})\cdot {\bf \nabla} \rho_o({\bf r})
\right) +
\nonumber
\\
+i \omega \sum_{k^\prime=1}^{n-1}
{\bf \nabla} \cdot  (\delta \rho_{k^\prime}({\bf r})
{\bf v}_{n-k^{\prime}}({\bf r}))
\label{rho_n}
\end{eqnarray}

The left hand
side of  equation (\ref{rho_n}) describes eigen-oscillations
of the electron density.  The electron density is
almost constant within the cluster but varies rapidly near
the cluster surface. Therefore, the terms proportional to
$\rho_o({\bf r})$ and ${\bf \nabla}\rho_o({\bf r})$
on the left hand side of  (\ref{rho_n})
determine the square of the frequency of the volume
and surface plasmon oscillations respectively.
The right hand side in (\ref{rho_n}) describes
a driving force acting on the eigen-
plasmon oscillations.

The set of non-linear equations (\ref{v_n}) and (\ref{rho_n})
must be solved iteratively.
It is clear from the form of the equations that the index $n$
corresponds to the order of perturbation theory on
the external field $\varphi({\bf r})$.

Indeed, for $n=1$ equation (\ref{rho_n}) reduces to
\begin{eqnarray}
\left(\omega^2 - \frac{4\pi e}{m} \rho_o({\bf r})\right)
\delta \rho_1({\bf r}) +
\frac{e}{m} {\bf \nabla}  \rho_o({\bf r}) \cdot
{\bf \nabla}
\int d{\bf r^\prime}
\frac{\delta \rho_1({\bf r^\prime})}{|{\bf r}-{\bf r^\prime}|}=
\nonumber
\\
=\frac{e}{m}
\left(\rho_o({\bf r})\varphi({\bf r}) k^2 -
{\bf \nabla} \varphi({\bf r})\cdot {\bf \nabla} \rho_o({\bf r})
\right)
\label{rho_n1}
\end{eqnarray}
and, for $n=2$, one derives
\begin{eqnarray}
\left((2\omega )^2 - \frac{4\pi e}{m} \rho_o({\bf r})\right)
\delta \rho_2({\bf r}) +
\frac{e}{m} {\bf \nabla}  \rho_o({\bf r}) \cdot
{\bf \nabla}
\int d{\bf r^\prime}
\frac{\delta \rho_2({\bf r^\prime})}{|{\bf r}-{\bf r^\prime}|}=
\nonumber
\\
= i \omega {\bf \nabla} \cdot  (\delta \rho_{1}({\bf r})
{\bf v}_{1}({\bf r}))
\label{rho_n_2}
\end{eqnarray}

Equations (\ref{rho_n1}) and (\ref{rho_n_2}) show that
the variation $\rho_1({\bf r^\prime})$ describes the
linear response of the electron subsystem to the
the external field $\varphi({\bf r})$, while
$\rho_2({\bf r^\prime})$ arises only in the second
order of perturbation theory on $\varphi({\bf r})$,
because $\rho_1({\bf r^\prime})\sim \varphi({\bf r})$
and ${\bf v}_{1}({\bf r}) \sim \varphi({\bf r})$.

Solving the set of equations
(\ref{v_n}) and (\ref{rho_n})
with  $\varphi({\bf r})$  describing
the dipole electron-photon interaction
up to the $n$-th order, one can calculate
the variation of electron density in the cluster caused
by the field of $n$ photons.

The set of equations (\ref{v_n}) and (\ref{rho_n})
is not confined in
its application to photons. It can also be used to describe
the dynamics of electron density under the action of
any kind  of external field, for example the electric field
of a charged projectile  colliding with the cluster.
Indeed, by considering the partial spherical harmonic
of the Fourier image of the Coulomb field of the projectile
particle, one can derive from (\ref{rho_n1}) the same
expression for the variation of the electron density
$\delta \rho_1({\bf r})$ as follows from
the purely electrodynamical perturbative
approach to the electron scattering problem \cite{ioni_cl}.

\subsection{Spherically symmetric case}

Equations (\ref{v_n}) and (\ref{rho_n})  are valid for
an arbitrary shape of the initial distribution
$\rho_{o}({\bf r})$. In the  spherical case, the angular
parts in  (\ref{v_n}) and (\ref{rho_n}) can be
separated. Thus, the cross
section for $n$-photon absorption can
in principle be extracted for arbitrarily large n,
although the calculations become more and more
tedious the higher $n$ is. Let us consider this
formalism in more detail.

In the case of the spherically symmetric initial
distribution, one can put $\rho_o({\bf r})= \rho_0(r)$.
This relationship allows one easily to exclude angular
variables from equation (\ref{rho_n}). Using this
relationship together with the partial expansion for
$\delta\rho_{n}({\bf r})$ and $\varphi({\bf r})$,
\begin{equation}
\delta\rho_{n}({\bf r})= \sum_{l=0}^{\infty}\sum_{m=-l}^{l}
\delta\rho_{l,m}^{n}(r) Y_{l,m}({\bf n}_r)
\label{partial_dens}
\end{equation}
\begin{equation}
\varphi({\bf r})= \sum_{l=0}^{\infty}\sum_{m=-l}^{l}
\varphi_{l,m}(r) Y_{l,m}({\bf n}_r),
\label{partial_varphi}
\end{equation}
one derives
\begin{eqnarray}
\left((\omega n)^2 - \frac{4\pi e}{m} \rho_o(r)\right)
\delta \rho_{l,m}^{n}(r) +
\frac{4\pi e \rho_o^{\prime}(r)}{m(2l+1)}
\int dr^\prime G_l(r,r^\prime)
\delta \rho^n_{l,m}(r^\prime)=
\nonumber
\\
=\frac{e}{m}\delta_{n,1}
\left(\rho_o(r)\varphi_{l,m}(r) k^2 -
\varphi_{l,m}^{\prime}(r)\rho_o^{\prime}(r)
\right) -
\frac{e}{m}
\sum_{j=1}^{n-1}
\frac{1}{n-j}
\int d\Omega_{{\bf n}_r}
Y_{l,m}^{*}({\bf n}_r) \times
\nonumber
\\
\times\Biggl\{
{\bf \nabla} \delta \rho_j({\bf r})\cdot {\bf \nabla}
\left(\delta_{n-j,1} \varphi({\bf r})
+
\int d{\bf r^{\prime}}
\frac{\delta \rho_{n-j}({\bf r^\prime})}{|{\bf r}-{\bf r^\prime}|}
\right)-
\nonumber\\
-\delta
\rho_j({\bf r})
(\delta_{n-j,1} \varphi({\bf r}) k^2+
4\pi \delta \rho_{n-j}({\bf r})
)
\Biggr\}
\label{rho_n_partial1}
\end{eqnarray}

Here,   $Y_{l,m}({\bf n}_r)$ is the sperical harmonic
corresponding to the angular momentum $l$ and the
projection of the angular momentum $m$.
When deriving (\ref{rho_n_partial1}), we have multiplied
both sides of equation
(\ref{rho_n}) by the sperical harmonic $Y_{l,m}^{*}({\bf n}_r)$
and then integrated over
$d\Omega_{{\bf n}_r}$. We also used the well known
expansion (see e.g. \cite{VMH})
\begin{equation}
\frac{1}{|{\bf r}-{\bf r^\prime}|}= \frac{1}{2l+1}
\sum_{l=0}^{\infty}\sum_{m=-l}^{l} B_l(r, r^\prime)
Y_{l,m}({\bf n}_r)Y_{l,m}^{*}({\bf n}_r^\prime)
\label{Coulomb_exp}
\end{equation}
where function $B_l(r, r^\prime)$ is defined as follows
\begin{equation}
B_l(r, r^\prime)=\frac{r^l}{r^{\prime l+1}} \Theta(r^\prime-r)+
\frac{r^{\prime l}}{r^{l+1}} \Theta(r-r^\prime)
\label{B-func}
\end{equation}
Here $\Theta(r^\prime-r)$ is the step function. In equation
(\ref{rho_n_partial1}) we have  introduced the function
$G_l(r, r^\prime)= r^2 \partial B_l(r, r^\prime)/\partial r $, which
is of the the form
\begin{equation}
G_l(r, r^\prime)=l\frac{r^{l-1}}{r^{\prime l-1}} \Theta(r^\prime-r)-
(l+1)\frac{r^{\prime l+2}}{r^{l+2}} \Theta(r-r^\prime)
\label{G-func}
\end{equation}
When deriving (\ref{rho_n_partial1}), we have also made
obvious transformations of the sum over $k^\prime$
on the right hand side of  equation (\ref{rho_n})
using (\ref{varphi_r})
and (\ref{v_n}). Note that  the sum over $j$ in
(\ref{rho_n_partial1}) still contains the integrals over the angular
variables. The integration over the angular variables in the sum
is straightforward, but somewhat cumbersome.
It is also clear that the  non-gradient terms in the sum
contain the integration of
the product of three spherical harmonics, if one
expands $\delta \rho_{n-j}({\bf r})$ and $\varphi({\bf r})$
according to  (\ref{partial_dens}) and (\ref{partial_varphi}),
which we denote as
\begin{equation}
I_1(l,m|l_1,m_1,|l_2,m_2)=
\int d\Omega_{{\bf n}_r}
Y_{l,m}^{*}({\bf n}_r) Y_{l_1,m_1}({\bf n}_r) Y_{l_2,m_2}({\bf n}_r)
\label{I1}
\end{equation}

The gradient terms in the sum contain the integration of
a spherical harmonic multiplied by the scalar product of
the two vector spherical harmonics:
\begin{eqnarray}
I_2(l,m|l_1,m_1,|l_2,m_2)=
\sqrt{l_1(l_1+1) l_2(l_2+1)}
\int d\Omega_{{\bf n}_r}\times
\nonumber
\\
Y_{l,m}^{*}({\bf n}_r) {\bf Y}^{(1)}_{l_1,m_1}({\bf n}_r)
\cdot{\bf Y}^{(1)}_{l_2,m_2}({\bf n}_r).
\label{I2}
\end{eqnarray}

This type of integral arises, when one
expresses the gradients of the potential and the density
according to  (see e.g. \cite{VMH})
\begin{eqnarray}
{\bf \nabla} \delta \rho^j_{l_1,m_1}(r) Y_{l_1,m_1}({\bf n}_r)=
\nonumber
\\
=\delta \rho^{j \prime}_{l_1,m_1}(r) {\bf Y}^{(-1)}_{l_1,m_1}({\bf n}_r)+
\sqrt{l_1(l_1+1)}\frac{1}{r} \delta \rho^{j }_{l_1,m_1}(r)
{\bf Y}^{(1)}_{l_1,m_1}({\bf n}_r)
\label{grad_rho}
\end{eqnarray}
\begin{eqnarray}
{\bf \nabla} \Phi^{n-j}_{l_2,m_2}(r) Y_{l_2,m_2}({\bf n}_r)=
\nonumber
\\
=\Phi^{(n-j) \prime}_{l_2,m_2}(r) {\bf Y}^{(-1)}_{l_2,m_2}({\bf n}_r)+
\sqrt{l_2(l_2+1)}\frac{1}{r} \Phi^{(n-j) }_{l_2,m_2}(r)
{\bf Y}^{(1)}_{l_2,m_2}({\bf n}_r)
\label{grad_phi}
\end{eqnarray}

Here ${\bf Y}^{(-1)}_{l,m}({\bf n}_r)$,
${\bf Y}^{(1)}_{l,m}({\bf n}_r)$ are, respectively, the longitudinal
and the transverse vector spherical harmonics,
the definition of which one can find in \cite{VMH}. We mention
some properties of these vector hamonics:
${\bf Y}^{(-1)}_{l,m}({\bf n}_r)= {\bf n_r} Y_{l,m}({\bf n}_r)$,
${\bf Y}^{(1)}_{l,m}({\bf n}_r)=
{\bf \nabla_\Omega}Y_{l,m}({\bf n}_r)/ \sqrt{l(l+1)}$
and
${\bf Y}^{(-1)}_{l,m}({\bf n}_r) \cdot {\bf Y}^{(1)}_{l,m}({\bf n}_r) =0$.

The potential $\Phi^{n-j}_{l_2,m_2}(r)$ in (\ref{grad_phi}) is
as follows
\begin{equation}
\Phi_{l_2,m_2}^{n-j}(r)=
\delta_{n-j,1} \varphi_{l_2,m_2}(r)+
\frac{4\pi}{(2l_2+1)} \int dr^{\prime}
r^{\prime 2} B_l(r,r^{\prime})
\delta \rho^{n-j}_{l_2,m_2}(r^\prime)
\label{Psi}
\end{equation}

Since the explicit expressions for
$I_1(l,m|l_1,m_1,|l_2,m_2)$ and
$I_2(l,m|l_1,m_1,|l_2,m_2)$
are somewhat lengthy, they are presented
in \ref{integrals}.

Using the formulae written above, one can easily
rewrite  (\ref{rho_n_partial1}) in the following
form
\begin{eqnarray}
\left((\omega n)^2 - \frac{4\pi e}{m} \rho_o(r)\right)
\delta \rho_{l,m}^{n}(r) +
\frac{4\pi e \rho_o^{\prime}(r)}{m(2l+1)}
\int dr^\prime G_l(r,r^\prime)
\delta \rho^n_{l,m}(r^\prime)=
\nonumber
\\
=\frac{e}{m}\delta_{n,1}
\left(\rho_o(r)\varphi_{l,m}(r) k^2 -
\varphi_{l,m}^{\prime}(r)\rho_o^{\prime}(r)
\right) -
\nonumber
\\
-\frac{e}{m}
\sum_{j=1}^{n-1}
\frac{1}{n-j}
\sum_{\stackrel{l_1,m_1}{l_2,m_2}} I_1(l,m|l_1,m_1,|l_2,m_2)
\times
\nonumber
\\
\times\left\{
\delta \rho^{j \prime}(r)
\Phi_{l_2,m_2}^{(n-j)\prime}(r)-
\delta \rho^{j}_{l_1,m_1}(r)
(\delta_{n-j,1} \varphi_{l_2,m_2}(r) k^2+
4\pi \delta \rho^{n-j}_{l_2,m_2}(r)) \right\}-
\nonumber
\\
-\frac{e}{m r^2}
\sum_{j=1}^{n-1}
\frac{1}{n-j}
\sum_{\stackrel{l_1,m_1}{l_2,m_2}} I_2(l,m|l_1,m_1,|l_2,m_2)
\delta \rho^j_{l_1,m_1}(r)
\Phi_{l_2,m_2}^{n-j}(r)
\label{rho_nlm}
\end{eqnarray}

Here and below, we assume that the summation over $l_1,m_1$ and $l_2,m_2$
is performed with the same limits as in (\ref{partial_dens})
and (\ref{partial_varphi}).

\subsection{Surface and volume plasmons}

We now analyse equation (\ref{rho_nlm}) and demonstrate
that it describes both surface and volume
plasmon oscillations. The surface and volume solutions of (\ref{rho_nlm})
can be separated from each other, if one assumes that the initial
distribution of electron density has the form
\begin{equation}
\rho_o(r) = \frac{Ne}{V}\Theta(R_v -r)
\label{rho_ini}
\end{equation}
Here $N$ is the total number of delocalized electrons in the cluster volume
$V= 4\pi R_v^3/3$.

In this case it is natural to look for the solution of (\ref{rho_nlm}) by
expressing it in the following form
\begin{equation}
\delta\rho^{(n)}_{l,m}(r)= \delta\rho^{s(n)}_{l,m}\delta (R_s-r)+
\delta\rho^{v(n)}_{l,m}(r)\Theta(R_v -r)
\label{delta_rho}
\end{equation}

In (\ref{rho_ini}) and (\ref{delta_rho}) we have introduced
the two radii $R_v$ and $R_s$ and assumed
that $ R_v < R_s =R$, but
$R_v \rightarrow R_s =R$, where $R$ is the cluster radius.
Such a relationship is necessary for the elimination of the
uncertainties, which arise in (\ref{delta_rho}) and the subsequent
formulae in the vicinity of the cluster radius.

\newpage

Substituting (\ref{rho_ini}) and (\ref{delta_rho}) in (\ref{rho_nlm}),
performing straighforward but lengthy calculations of the integrals,
one derives the following equation
\begin{eqnarray}
\left((\omega n)^2 - \omega_p^2  \right)
\delta \rho_{l,m}^{v(n)}(r) \Theta(R -r) +
\nonumber
\\
+\left((\omega n)^2 - \omega_l^2  \right)
\delta \rho_{l,m}^{s(n)} \delta(R -r) =
\nonumber
\\
=- \frac{4\pi e N (l+1)}{m (2l+1) V R^{l+2}} \delta(R-r)
\int_0^R dr^\prime  r^{\prime l+2} \delta \rho^{v(n)}_{l,m}(r^\prime)+
\nonumber
\\
+\frac{Ne^2 }{mV}\delta_{n,1}
\varphi_{l,m}(r) k^2 \Theta(R-r) +
\frac{Ne^2 }{mV} \delta_{n,1}
\varphi_{l,m}^{\prime}(R) \delta(R-r)+
\nonumber
\\
+\frac{e}{m} \delta(R-r)
\sum_{j=1}^{n-1}
\frac{1}{n-j}
\sum_{\stackrel{l_1,m_1}{l_2,m_2}} I_1(l,m|l_1,m_1,|l_2,m_2)
\times
\nonumber
\\
\times\Biggl\{
\delta \rho^{s(j)}_{l_1,m_1}
\left(\delta_{n-j,1}
\varphi_{l_2,m_2}^{\prime\prime}(R)+
\frac{4\pi}{(2l_2+1)}
\int_0^R dr^\prime  D_{l_2}(R,r^{\prime})
\delta\rho^{v(n-j)}_{l_2,m_2}(r^\prime)\right)+
\nonumber
\\
+\delta\rho^{v(j)}_{l_1,m_1}(R)
\left(\delta_{n-j,1} \varphi^{\prime}_{l_2,m_2}(R) +
\frac{4\pi}{(2l_2+1)}
\int_0^R dr^\prime  G_{l_2}(R,r^{\prime})
\delta\rho^{v(n-j)}_{l_2,m_2}(r^\prime)\right)+
\nonumber
\\
+\delta_{n-j,1}k^2
\delta\rho^{s(j)}_{l_1,m_1}\varphi_{l_2,m_2}(R) \Biggr\}-
\nonumber
\\
-\frac{e}{m} \Theta(R-r)
\sum_{j=1}^{n-1}
\frac{1}{n-j}
\sum_{\stackrel{l_1,m_1}{l_2,m_2}} I_1(l,m|l_1,m_1,|l_2,m_2)
\times
\nonumber
\\
\times\Biggl\{
\delta \rho^{v(j)\prime}_{l_1,m_1}(r)
\left(\delta_{n-j,1}
\varphi_{l_2,m_2}^{\prime}(r)+
\frac{4\pi}{(2l_2+1)}
\int_0^R dr^\prime  G_{l_2}(r,r^{\prime})
\delta\rho^{v(n-j)}_{l_2,m_2}(r^\prime)\right)-
\nonumber
\\
-\delta_{n-j,1}k^2\delta\rho^{v(j)}_{l_2,m_2}(r)\varphi_{l_2,m_2}(r)-
4\pi \delta\rho^{v(j)}_{l_2,m_2}(r) \delta\rho^{v(j)}_{l_2,m_2}(r)
\Biggr\}-
\nonumber
\\
-\frac{e}{m R^2 } \delta(R-r)
\sum_{j=1}^{n-1}
\frac{1}{n-j}
\sum_{\stackrel{l_1,m_1}{l_2,m_2}} I_2(l,m|l_1,m_1,|l_2,m_2)
\times
\nonumber
\\
\times
\delta \rho^{s(j)}_{l_1,m_1}
\left(\delta_{n-j,1}
\varphi_{l_2,m_2}(R)+
\frac{4\pi}{(2l_2+1)}
\int_0^R dr^\prime r^{\prime 2} B_{l_2}(R,r^{\prime})
\delta\rho^{v(n-j)}_{l_2,m_2}(r^\prime)\right)-
\nonumber
\\
-\frac{e}{m r^2 } \Theta(R-r)
\sum_{j=1}^{n-1}
\frac{1}{n-j}
\sum_{\stackrel{l_1,m_1}{l_2,m_2}} I_2(l,m|l_1,m_1,|l_2,m_2)
\times
\nonumber
\\
\times
\delta \rho^{v(j)}_{l_1,m_1}(r)
\left(\delta_{n-j,1}
\varphi_{l_2,m_2}(r)+
\frac{4\pi}{(2l_2+1)}
\int_0^R dr^\prime r^{\prime 2} B_{l_2}(r,r^{\prime})
\delta\rho^{v(n-j)}_{l_2,m_2}(r^\prime)\right)
\label{rho_nlm_fin}
\end{eqnarray}

When deriving (\ref{rho_nlm_fin}), we have used the fact that
$\Theta^\prime(R-r)= -\delta(R-r)$. Also we have introduced
the function
\begin{equation}
D_l(r, r^\prime)=l_2(l_2-1)\frac{r^{l_2-2}}{r^{\prime l_2-1}} \Theta(r^\prime-r)+
(l_2+1)(l_2+2)\frac{r^{\prime l_2+2}}{r^{l+3}} \Theta(r-r^\prime)
\label{D}
\end{equation}

The left hand
side of  equation (\ref{rho_nlm_fin}) describes volume and surface
eigen-oscillations of the electron density characterised by
the angular momentum $l$.
The surface plasmon resonance frequency $\omega_l$ is
the same as in (\ref{omega_pl}).
The volume plasmon resonance frequency is equal to
\begin{equation}
\omega_p=\sqrt{\frac{4\pi e^2 N}{m V}}
\label{vol_freq}
\end{equation}
In equation (\ref{rho_nlm_fin}), $\omega_p$ appears in expressions involving
$l$, but is nevertheless is independent of $l$, as one sees in equation
(\ref{vol_freq}).  The physical reason for this is that the volume plasmon
oscillation is degenerate with $l$.
The right hand side of (\ref{rho_nlm_fin}) describes
a driving force acting on the eigen-plasmon oscillations.

Surface and volume terms on the right hand side
of equation (\ref{rho_nlm_fin}) have not been regrouped, in order to
stress their correspondencce with terms in
(\ref{rho_nlm}). It is  seen from (\ref{rho_nlm_fin}) that equations
for the volume and surface  plasmon oscillations can be
separated and will then read as follows:
\begin{eqnarray}
\left((\omega n)^2 - \omega_p^2 \right)
\delta \rho_{l,m}^{v(n)}(r)=
\frac{Ne^2 }{mV}\delta_{n,1}
\varphi_{l,m}(r) k^2 -
\nonumber
\\
-\frac{e}{m}
\sum_{j=1}^{n-1}
\frac{1}{n-j}
\sum_{\stackrel{l_1,m_1}{l_2,m_2}} I_1(l,m|l_1,m_1,|l_2,m_2)
\times
\nonumber
\\
\times\Biggl\{
\delta \rho^{v(j)\prime}_{l_1,m_1}(r)
\left(\delta_{n-j,1}
\varphi_{l_2,m_2}^{\prime}(r)+
\frac{4\pi}{(2l_2+1)}
\int_0^R dr^\prime  G_{l_2}(r,r^{\prime})
\delta\rho^{v(n-j)}_{l_2,m_2}(r^\prime)\right)-
\nonumber
\\
-\delta_{n-j,1}k^2\delta\rho^{v(j)}_{l_2,m_2}(r)\varphi_{l_2,m_2}(r)-
4\pi \delta\rho^{v(j)}_{l_2,m_2}(r) \delta\rho^{v(j)}_{l_2,m_2}(r)
\Biggr\}-
\nonumber
\\
-\frac{e}{m r^2 }
\sum_{j=1}^{n-1}
\frac{1}{n-j}
\sum_{\stackrel{l_1,m_1}{l_2,m_2}} I_2(l,m|l_1,m_1,|l_2,m_2)
\times
\nonumber
\\
\times
\delta \rho^{v(j)}_{l_1,m_1}(r)
\left(\delta_{n-j,1}
\varphi_{l_2,m_2}(r)+
\frac{4\pi}{(2l_2+1)}
\int_0^R dr^\prime r^{\prime 2} B_{l_2}(r,r^{\prime})
\delta\rho^{v(n-j)}_{l_2,m_2}(r^\prime)\right)
\label{rho_nlm_vol}
\end{eqnarray}

\begin{eqnarray}
\left((\omega n)^2 - \omega_l^2  \right)
\delta \rho_{l,m}^{s(n)}  =
\frac{Ne^2 }{mV} \delta_{n,1}
\varphi_{l,m}^{\prime}(R)-
\nonumber
\\
- \frac{4\pi e N (l+1)}{m (2l+1) V R^{l+2}}
\int_0^R dr^\prime  r^{\prime l+2} \delta \rho^{v(n)}_{l,m}(r^\prime)+
\nonumber
\\
+\frac{e}{m}
\sum_{j=1}^{n-1}
\frac{1}{n-j}
\sum_{\stackrel{l_1,m_1}{l_2,m_2}} I_1(l,m|l_1,m_1,|l_2,m_2)
\times
\nonumber
\\
\times\Biggl\{
\delta \rho^{s(j)}_{l_1,m_1}
\left(\delta_{n-j,1}
\varphi_{l_2,m_2}^{\prime\prime}(R)+
\frac{4\pi}{(2l_2+1)}
\int_0^R dr^\prime  D_{l_2}(R,r^{\prime})
\delta\rho^{v(n-j)}_{l_2,m_2}(r^\prime)\right)+
\nonumber
\\
+\delta\rho^{v(j)}_{l_1,m_1}(R)
\left(\delta_{n-j,1} \varphi^{\prime}_{l_2,m_2}(R) +
\frac{4\pi}{(2l_2+1)}
\int_0^R dr^\prime  G_{l_2}(R,r^{\prime})
\delta\rho^{v(n-j)}_{l_2,m_2}(r^\prime)\right)+
\nonumber
\\
+\delta_{n-j,1}k^2
\delta\rho^{s(j)}_{l_1,m_1}\varphi_{l_2,m_2}(R) \Biggr\}-
\nonumber
\\
-\frac{e}{m R^2 }
\sum_{j=1}^{n-1}
\frac{1}{n-j}
\sum_{\stackrel{l_1,m_1}{l_2,m_2}} I_2(l,m|l_1,m_1,|l_2,m_2)
\times
\nonumber
\\
\times
\delta \rho^{s(j)}_{l_1,m_1}
\left(\delta_{n-j,1}
\varphi_{l_2,m_2}(R)+
\frac{4\pi}{(2l_2+1)}
\int_0^R dr^\prime r^{\prime 2} B_{l_2}(R,r^{\prime})
\delta\rho^{v(n-j)}_{l_2,m_2}(r^\prime)\right)
\label{rho_nlm_surf}
\end{eqnarray}

The set of non-linear equations (\ref{rho_nlm_vol}) and (\ref{rho_nlm_surf})
must be solved iteratively starting from $n=1$.
It is clear from the form of these equations that the index $n$
corresponds to the order of perturbation theory on
the external field $\varphi({\bf r})$.
It is also seen that the chain of equations for the volume plasmon
oscillations is self-sufficient, while the solutions
for the surface oscillations also depend on those for volume, which is
physically clear, because the electric field caused by volume oscillations
of electron density must influence surface oscillations.

From (\ref{rho_nlm_vol}) and (\ref{rho_nlm_surf}),
one derives the following solutions for $n=1$:
\begin{equation}
\delta \rho_{l,m}^{v(1)}(r)=
\frac{\omega_p^2 }{4\pi}
\frac{k^2\varphi_{l,m}(r)}{\omega^2 - \omega_p^2 }
\label{rho_nlm_vol1}
\end{equation}
\begin{equation}
\delta \rho_{l,m}^{s(1)}=
\frac{\omega_p^2 }{4\pi}
\frac{1}{\omega^2 - \omega_l^2 }
\left\{
\varphi_{l,m}^{\prime}(R)-
\frac{4\pi (l+1)}{(2l+1)}
\int_0^R dr^\prime  \frac{r^{\prime l+2}}{R^{l+2}}
\delta \rho^{v(1)}_{lm}(r^\prime)
\right\}
\label{rho_nlm_surf1}
\end{equation}

These equations show that, if the external field (\ref{partial_varphi}) is
characterized by a certain angular momentum $l_o$ and its projection
$m_o$, i.e.
$\varphi_{l,m}(r)=\delta_{l,l_o} \delta_{m,m_o}\varphi_{l_o,m_o}(r)$,
then
$\delta \rho_{l,m}^{s(1)}=
\delta_{l,l_o} \delta_{m,m_o} \delta \rho_{l_o,m_o}^{s(1)} $
and $\delta \rho^{v(1)}_{lm}(r)=
\delta_{l,l_o} \delta_{m,m_o} \delta \rho^{v(1)}_{l_o,m_o}(r)$.
Assuming these dependencies,
the solutions of (\ref{rho_nlm_vol}) and (\ref{rho_nlm_surf}) for
$n=2$ read as:

\begin{eqnarray}
\delta \rho_{l,m}^{v(2)}(r)=
-\frac{e}{m ((2\omega)^2 - \omega_p^2 )}
I_1(l,m|l_o,m_o,|l_o,m_o)
\times
\nonumber
\\
\times\left\{
\delta \rho^{v(1)\prime}_{l_o,m_o}(r)
\left(
\varphi_{l_o,m_o}^{\prime}(r)+
\frac{4\pi}{(2l_2+1)}
\int_0^R dr^\prime  G_{l_o}(r,r^{\prime})
\delta\rho^{v(1)}_{l_o,m_o}(r^\prime)\right)-
\right.
\nonumber
\\
\left.
-k^2\delta\rho^{v(1)}_{l_o,m_o}(r)\varphi_{l_0,m_o}(r)-
4\pi \delta\rho^{v(1)}_{l_o,m_o}(r) \delta\rho^{v(1)}_{l_o,m_o}(r)
\right\}-
\nonumber
\\
-\frac{e}{m r^2 ((2\omega)^2 - \omega_p^2 ) }
I_2(l,m|l_o,m_o,|l_o,m_o)
\times
\nonumber
\\
\times
\delta \rho^{v(1)}_{l_o,m_o}(r)
\left(
\varphi_{l_o,m_o}(r)+
\frac{4\pi}{(2l_o+1)}
\int_0^R dr^\prime r^{\prime 2} B_{l_o}(r,r^{\prime})
\delta\rho^{v(1)}_{l_o,m_o}(r^\prime)\right)
\label{rho_nlm_vol2}
\end{eqnarray}

\begin{eqnarray}
\delta \rho_{l,m}^{s(2)}  =
- \frac{4\pi e N (l+1)}{m (2l+1) V R^{l+2} ((2\omega )^2 - \omega_l^2  )}
\int_0^R dr^\prime  r^{\prime l+2} \delta \rho^{v(1)}_{l_o,m_o}(r^\prime)+
\nonumber
\\
+\frac{e}{m ((2\omega )^2 - \omega_l^2  )}
I_1(l,m|l_o,m_o,|l_o,m_o)
\times
\nonumber
\\
\times\left\{
\delta \rho^{s(1)}_{l_o,m_o}
\left(
\varphi_{l_o,m_o}^{\prime\prime}(R)+
\frac{4\pi}{(2l_o+1)}
\int_0^R dr^\prime  D_{l_o}(R,r^{\prime})
\delta\rho^{v(1)}_{l_o,m_o}(r^\prime)\right)+
\right.
\nonumber
\\
\left.
+\delta\rho^{v(1)}_{l_o,m_o}(R)
\left(\varphi^{\prime}_{l_o,m_o}(R) +
\frac{4\pi}{(2l_o+1)}
\int_0^R dr^\prime  G_{l_o}(R,r^{\prime})
\delta\rho^{v(1)}_{l_o,m_o}(r^\prime)\right)+
\right.
\nonumber
\\
\left.
+k^2
\delta\rho^{s(1)}_{l_o,m_o}\varphi_{l_o,m_o}(R) \right\}-
\nonumber
\\
-\frac{e}{m R^2 ((2\omega )^2 - \omega_l^2  ) }
I_2(l,m|l_o,m_o,|l_o,m_o)
\times
\nonumber
\\
\times
\delta \rho^{s(1)}_{l_o,m_o}
\left(
\varphi_{l_o,m_o}(R)+
\frac{4\pi}{(2l_o+1)}
\int_0^R dr^\prime r^{\prime 2} B_{l_o}(R,r^{\prime})
\delta\rho^{v(1)}_{l_o,m_o}(r^\prime)\right)
\label{rho_nlm_surf2}
\end{eqnarray}

By performing similar transformations, one can find the solutions
$\delta \rho_{lm}^{v(n)}(r)$ and $\delta \rho_{lm}^{s(n)}$
for arbitrarily large $n$, although the formulae become more
and more tedious the larger $n$ becomes. These formulae demonstrate
that, in the higher orders of perturbation theory, plasmon
resonances with angular momenta larger than the angular
momentum of the external field can be excited. Indeed, the selection
rules for the integrals  $I_1(l,m|l_o,m_o,|l_o,m_o)$ and
$I_2(l,m|l_o,m_o,|l_o,m_o)$ (see \ref{integrals}) show that
the angular momentum in $\delta \rho_{lm}^{v(2)}(r)$  and
$\delta \rho_{lm}^{s(2)}$ can be  twice as large as $l_o$.
Equations (\ref{rho_nlm_vol}- \ref{rho_nlm_surf2})
also demonstrate that the plasmon resonances in
$\delta \rho_{lm}^{v(n)}(r)$ and $\delta \rho_{lm}^{s(n)}$
arise when  $\omega= \omega_p/n$
and $\omega= \omega_l/n$ respectively. These equations indicate
a significant shift of the plasmon resonance profiles towards lower frequencies
in the highest orders of perturbation theory.

These results have a simple physical explanation. Absorption
of several quanta of the external field (photons) by the cluster leads to
the excitation of non-dipole plasmon oscillations
of the electron density.

\subsection{Fast electron-cluster collisions}

Equations (\ref{rho_nlm_vol}- \ref{rho_nlm_surf2})
can be used for the analysis of the balance between
the surface and volume plasmon oscillations in the
cluster. We demonstrate this for the example
of fast electron scattering on a metal cluster.
In this case, the external field of the projectile
electron can be characterized by the Fourier component
of the Coulomb potential
\begin{equation}
\varphi(r) =\frac{4\pi}{q^2} e^{i{\bf q}{\bf r}},
\label{el_pot}
\end{equation}
where ${\bf q}={\bf p}- {\bf p}^\prime$ is the transferred
momentum of the scattered electron.

The partial expansion of this potential reads as:
\begin{equation}
\varphi(r) =4\pi\sum_{l=0}^{\infty} \sum_{m=-}^{m=l}
i^l\varphi_{l,m}(r) Y_{l,m}^{*}({\bf n}_q),
\label{el_pot_par_exp}
\end{equation}
where the partial component of the potential $\varphi_{l,m}(r)$
is equal to:
\begin{equation}
\varphi_{l,m}(r) = \frac{4\pi}{q^2} j_l(q r) Y_{l,m}({\bf n}_r),
\label{el_pot_parial}
\end{equation}
and $j_l(q r)$ is the spherical Bessel function
(for definition see e.g.\cite{LL3}).

The form and properties of $\varphi_{l,m}(r)$ - (\ref{el_pot_parial})
are exactly the same as  assumed  in (\ref{varphi_r})
and (\ref{partial_varphi}).
Therefore, from (\ref{rho_nlm_vol1}) and (\ref{rho_nlm_surf1}),
one can immediately derive:
\begin{equation}
\delta \rho_{l,m}^{v(1)}(r)=
\frac{ \omega_p j_{l}(qr)}{\omega^2 - \omega_p^2 }
\label{el_rho_nlm_vol1}
\end{equation}
\noindent
and
\begin{equation}
\delta \rho_{l,m}^{s(1)}=
\frac{\omega_p}{q^2(\omega^2 - \omega_l^2) }
\left\{
j_{l}^{\prime}(qR)-
\frac{q^2 \omega_p (l+1)}{ (\omega^2 - \omega_p^2) (2l+1)}
\int_0^R dr^\prime  \frac{r^{\prime l+2}}{R^{l+2}}
j_{l}(qr^\prime)
\right\}
\label{el_rho_nlm_surf1}
\end{equation}

In the case of inelastic electron  scattering, $\omega$ has
the meaning of the transferred energy in the collision,
$\omega=\Delta\varepsilon =\varepsilon- \varepsilon^\prime$.
Calculating the integral in (\ref{el_rho_nlm_surf1})
with the use of the well known properties of  spherical
Bessel functions (see e.g. \cite{LL3}), one derives
\begin{eqnarray}
\delta \rho_{l,m}^{s(1)}
&=&
\frac{(2l+1)\omega _l^2}{\omega^2-\omega _l^2}
\frac{j_l(qR)}{q^2R}-
\frac{\omega _p^2}{\omega^2-\omega _p^2}
j_{l+1}(qR)
\label{el_rho_nlm_fin}
\end{eqnarray}

Expresssions (\ref{el_rho_nlm_vol1}) and (\ref{el_rho_nlm_fin})
coincide with those calculated in \cite{ioni_cl,Lushnikov74}
in the plasmon resonance approximation by purely
electrodynamic means, as the response of a
dielectric sphere, having
dielectric permeability $\epsilon =1-
\theta (R-r) \omega ^2/\omega _p^2$.

From (\ref{el_rho_nlm_vol1}) and (\ref{el_rho_nlm_fin}),
one can easily elaborate the electron inelastic
scattering cross section in the plasmon resonance  approximation,
using the method described in \cite{ioni_cl}:
\begin{eqnarray}
\frac{d^2\sigma }{d\varepsilon ^{\prime }d\Omega } &=&\frac{4p^{\prime }R}{
\pi pq^4}\sum_l(2l+1)^2j_l^2(qR)\frac{\omega _l^2\Delta \varepsilon
\Gamma^s_{l}}{({\Delta \varepsilon }^2-{\omega _l}^2)^2+
{\Delta \varepsilon }^2{\Gamma}_{l}^{s2}}+  \label{eq:8} \\
&+&\frac{2p^{\prime }R^3}{\pi pq^2}
\sum_l(2l+1)
\frac{\omega _p^2\Delta \varepsilon
\Gamma^v_{l}}{({\Delta \varepsilon }^2-
{\ \omega _p}^2)^2+\Delta \varepsilon ^2\Gamma_{l}^{v2}} \times \nonumber\\
&\times&
\left(j_l^2(qR)-j_{l+1}(qR)j_{l-1}(qR)-
\frac 2{qR}j_{l+1}(qR)j_l(qR) \right)
\nonumber
\end{eqnarray}

This cross section is totally determined by collective
electron excitations in the cluster.
The first and the second terms
in (\ref{eq:8}) describe the contributions of
the surface and the volume plasmon excitations
respectively. In (\ref{eq:8}), we have also indroduced
widths, $\Gamma^s_{l}$ and $\Gamma^v_{l}$,
of the surface and volume plasmon resonances.
They originate from the Landau damping mechanism
of the plasmon excitations. For their determination
we refer to the recent paper \cite{ioni_cl}.

\subsection{Multiphoton absorption}

Next, we apply equations (\ref{rho_nlm_vol}) and (\ref{rho_nlm_surf})
to the description of the multiphoton absorption process.
In this paper, we focus our consideration on the analysis of
plasmon excitations. If surface or volume plasmon resonances
are excited by photons, i.e. $\omega\sim\omega_p$,
then it is easy to check that the following condition is fulfilled
$\omega R/c \sim \omega_p R/c\ll 1$, where
$c$ is the velocity of light. This condition implies the validity
of the dipole approximation.

In the dipole approximation, one can neglect the momentum
of the photon and put  $k=0$. In this case,  equations
(\ref{rho_nlm_vol}) and (\ref{rho_nlm_surf}) are simplified
dramaticaly. Indeed, from (\ref{rho_nlm_vol}), one derives
\begin{equation}
\delta \rho_{l,m}^{v(n)}(r)= 0
\label{dipole_v}
\end{equation}

This result also simplifies  equation (\ref{rho_nlm_surf}) significantly for
$\delta \rho_{lm}^{s(n)}$. After some trivial transformations
it reduces to:
\begin{eqnarray}
\left((\omega n)^2 - \omega_l^2  \right)
\delta \rho_{l,m}^{s(n)}  =
\frac{Ne^2 }{mV} \delta_{n,1}
\varphi_{lm}^{\prime}(R)+
\nonumber
\\
+\frac{e}{m}
\sum_{j=1}^{n-1}
\frac{\delta_{n-j,1}}{n-j}
\sum_{\stackrel{l_1,m_1}{l_2,m_2}} I_1(l,m|l_1,m_1,|l_2,m_2)
\delta \rho^{s(j)}_{l_1,m_1}
\varphi_{l_2,m_2}^{\prime\prime}(R)-
\nonumber
\\
-\frac{e}{m R^2 }
\sum_{j=1}^{n-1}
\frac{\delta_{n-j,1}}{n-j}
\sum_{\stackrel{l_1,m_1}{l_2,m_2}} I_2(l,m|l_1,m_1,|l_2,m_2)
\delta \rho^{s(j)}_{l_1,m_1}
\varphi_{l_2,m_2}(R)
\label{dipole_s}
\end{eqnarray}

The partial  component of the linearely polarized
dipole photon field is equal to
\begin{equation}
\varphi_{l,m}(r)= - \sqrt{\frac{4\pi}{3}}  E r
\delta_{l,1} \delta_{m,0}
\label{dipole-int}
\end{equation}
Here $E= \sqrt{2\pi\hbar\omega/V_o}$ is the strength of
the photon's electric field and $V_o$ is the normalization volume of
the photon mode.
Substituting (\ref{dipole-int}) into (\ref{rho_nlm_surf}),
one derives
\begin{eqnarray}
\left((\omega n)^2 - \omega_l^2  \right)
\delta \rho_{l,m}^{s(n)}  =
-\sqrt{\frac{4\pi}{3}} \frac{Ne^2 E }{mV} \delta_{n,1}
\delta_{l,1} \delta_{m,0}+
\nonumber
\\
\\
+ \sqrt{\frac{4\pi}{3}}\frac{e E}{m R }
\sum_{l_1,m_1} I_2(l,m|l_1,m_1,|1,0)
\delta \rho^{s(n-1)}_{l_1,m_1}
\label{rho_nlm_surf_dip}
\end{eqnarray}

This equation should be solved iteratively
starting from $n=1$. For $n=1$, the single non-trivial
solution, $\delta \rho_{1,0}^{s(1)}$, reads as
\begin{equation}
\delta \rho_{1,0}^{s(1)}  =
-\sqrt{\frac{4\pi}{3}} \frac{Ne^2 E }{mV (\omega^2 - \omega_1^2) }
\label{rho_110_surf}
\end{equation}

Then, for $n=2$, the solution of (\ref{rho_nlm_surf_dip}) is
of the form
\begin{eqnarray}
\delta \rho_{l,m}^{s(2)}  =
\sqrt{\frac{4\pi}{3}}
\frac{e E \delta \rho^{s(1)}_{1,0}}{m R ((2\omega )^2 - \omega_l^2) }
I_2(l,m|1,0|1,0)
\label{rho_2lm_surf}
\end{eqnarray}

The selection rules for $I_2(l,m|1,0|1,0)$ (see \ref{integrals}) show
that this integral does not vanish, when
$l=0$ and $m=0$ or $l=2$ and $m=0$.
Therefore, for $n=3$ from (\ref{rho_nlm_surf_dip}), one derives
\begin{eqnarray}
\delta \rho_{l,m}^{s(3)}  =
\sqrt{\frac{4\pi}{3}}
\frac{e E }{m R((3\omega)^2 - \omega_l^2) }
\nonumber
\\
\left\{
I_2(l,m|0,0|1,0)\delta \rho^{s(2)}_{0,0}+
I_2(l,m|2,0|1,0)\delta \rho^{s(2)}_{2,0}\right\}
\label{rho_3lm_surf}
\end{eqnarray}

In (\ref{rho_3lm_surf}), only the
second term in brackets gives a non-zero contribution,
since  $I_2(l,m|0,0|1,0)=0$ (see \ref{integrals}).
Substituting  $\delta\rho^{s(2)}_{0,0}$ and
$\delta\rho^{s(2)}_{2,0}$ from (\ref{rho_2lm_surf}) and
$\delta \rho_{1,0}^{s(1)}$ from (\ref{rho_110_surf})
in (\ref{rho_3lm_surf})
and
using the explicit expressions for the angular integrals given
in  \ref{integrals}, one obtains
\begin{eqnarray}
&&
\delta \rho_{0,0}^{s(2)}=
-\frac{\pi^{1/2}}{3m^2 R V}
\cdot
\frac{Ne^3E^2}{ \omega^2
(\omega^2- \omega_1^2) }
\nonumber
\\
&&
\delta \rho_{2,0}^{s(2)}=
-\frac{4\pi^{1/2}}{3\sqrt{5}m^2 R V}
\cdot
\frac{Ne^3E^2}{
(\omega^2- \omega_1^2)((2\omega)^2- \omega_2^2) }
\nonumber
\\
&&
\delta \rho_{1,0}^{s(3)}=
-\left(\frac{64\pi}{3}\right)^{1/2}
\frac{16+3\sqrt{5}}{75m^3 R^2 V}
\cdot
\frac{N e^4 E^3}{
(\omega^2- \omega_1^2)((2 \omega)^2- \omega_2^2 )
((3 \omega)^2 -\omega_1^2 )}
\nonumber
\\
&&
\delta \rho_{3,0}^{s(3)}= 
\left(\frac{4\pi}{7}\right)^{1/2}
\frac{12 (2+\sqrt{5})}{75m^3 R^2 V}
\cdot
\frac{N e^4 E^3}{
(\omega^2- \omega_1^2)((2 \omega)^2- \omega_2^2 )
((3 \omega)^2 -\omega_3^2 )}
\label{rho_3}
\end{eqnarray}

\section{Induced multipole moments in the cluster}
\label{q_mom}

Let us now calculate the multipole
moments of the cluster induced by an external radiation
field on the basis of the model developed in the previous
section, and analyse their plasmon resonance structure.

The induced multipole moment of the cluster is equal to
\begin{equation}
Q_{l,m}=\sqrt{\frac{4\pi}{2l+1}} \int dV r^l Y_{l,m}({\bf n}_r)
\delta \rho({\bf r})
\label{mult_mom_def}
\end{equation}
\noindent
where the variation of electron density $\delta \rho({\bf r})$
is determined in (\ref{partial_dens}) and (\ref{delta_rho}).
Subsituting  (\ref{partial_dens}) and
(\ref{delta_rho}) in (\ref{mult_mom_def})
and putting
$\delta\rho^{v(n)}_{l,m}(r)=0$ for any $n$ in the dipole approximation as
follows from (\ref{dipole_v}), one derives
\begin{equation}
Q_{l,m}^{(n)}=\sqrt{\frac{4\pi}{2l+1}}  R^{l+2} \delta\rho^{s(n)}_{l,m}
\label{mult_mom}
\end{equation}

Substituting in (\ref{mult_mom})  $\delta\rho^{s(n)}_{1,0}$
from (\ref{rho_110_surf}), one obtains
the expression for
the dipole moment of the cluster induced
in the single-photon absorption process
\begin{equation}
D^{(1)}(\omega)=Q_{1,0}^{(1)}=
-\frac{Ne^2 E}{m
(\omega^2 -\omega_1^2 +\i \omega\Gamma_1)
}
\label{dip_mom_1}
\end{equation}

The explicit expressions
for the partial electron density variations $\delta\rho^{s(n)}_{l,m}$
entering
(\ref{mult_mom}) for  $n=2$ (two photon case) and $n=3$
(three photon case) are given in (\ref{rho_3}).
Subsituting
the partial electron density variations 
$\delta\rho^{s(2)}_{0,0}$ and $\delta\rho^{s(2)}_{2,0}$
from (\ref{rho_3}) into  (\ref{mult_mom}),
one derives the expression for
the monopole and quadrupole moment of the cluster induced
in the two photon regime
\begin{eqnarray}
M^{(2)}(\omega)=Q_{0,0}^{(2)}= -\frac{1}{2 m^2 R^2}
\frac{Ne^3 E^2}{ \omega^2
(\omega^2 -\omega_1^2 +\i \omega\Gamma_1)
}
\nonumber
\\
Q^{(2)}(\omega)=Q_{2,0}^{(2)}= -\frac{2}{5}
\frac{Ne^3 E^2}{m^2
(\omega^2 -\omega_1^2 +\i \omega\Gamma_1)
((2\omega)^2 -\omega_2^2 + i2\omega\Gamma_2)
}
\label{quad_mom}
\end{eqnarray}
Here, we have introduced
the plasmon resonance
widths $\Gamma_1$ and $\Gamma_2$ which take into account
Landau damping of the dipole and quadrupole surface plasmon
resonances. They must be
determined separately, e.g. by an {\it ab initio} approach
(see \cite{ioni_cl}).

By absorbing three photons one can induce dipole and octupole moments in the
cluster.  Substituting $\delta \rho_{1,0}^{s(3)}$ from (\ref{rho_3})
into (\ref{mult_mom}), one derives the expression for the
induced dipole moment
\begin{eqnarray}
D^{(3)}(\omega)&=&Q_{1,0}^{(3)}= - \frac{4(16+ 3\sqrt{5})}{75m^3 R^2}\times
\nonumber
\\
&\times&
\frac{Ne^4 E^3}{(\omega^2 -\omega_1^2 +\i \omega\Gamma_1)
((2\omega)^2 -\omega_2^2 + i2\omega\Gamma_2)
((3\omega)^2 -\omega_1^2 +\i 3\omega\Gamma_1)}
\label{dip_mom_3}
\end{eqnarray}

The expression for the octupole moment induced by 3 photons following
from (\ref{rho_3}) and (\ref{mult_mom}) reads as
\begin{eqnarray}
O^{(3)} (\omega)&=&Q_{3,0}^{(3)}= \frac{12(2+ \sqrt{5})}{175m^3 }\times
\nonumber
\\
&\times&
\frac{Ne^4 E^3}{(\omega^2 -\omega_1^2 +\i \omega\Gamma_1)
((2\omega)^2 -\omega_2^2 + i2\omega\Gamma_2)
((3\omega)^2 -\omega_3^2 +\i 3\omega\Gamma_3)}
\label{oct_mom}
\end{eqnarray}

Here, we have also introduced
the octupole plasmon resonance width $\Gamma_3$.

Expressions (\ref{dip_mom_1}-\ref{oct_mom}) demonstrate that the
multipole moments induced in the cluster during multiphoton absorption
processes possess a prominent plasmon resonance structure.
The nature
of these resonances is the same as occurs in the mutiphoton
absorption cross sections discussed in section \ref{m_phot}.

The connection between $D^{(1)}(\omega)$ from (\ref{dip_mom_1})
and  the cross section $\sigma_1$ found in (\ref{photo_1})
is straightforward
\begin{equation}
\sigma_1=\frac{4\pi \omega}{c E} {\it Im} D^{(1)}(\omega)
\label{sigma1_D1}
\end{equation}

In the multiphoton regime, the connection between the induced
multipole moments of the cluster and the multiphoton absorption
cross section becomes more complex, which is apparent from the
the classical nature of the expressions (\ref{quad_mom}-\ref{oct_mom})
and the explicit dependence of the multiphoton absorption
cross sections on Planck's constant. The discussion of this
interesting relationship is however beyond the scope of the present paper.

\section{Conclusion}
\label{conclu}

In this paper, we have developed a formalism which
allows one to calculate the cross section
for multiphoton absorption in the plasmon
resonance approximation. We have demonstrated that
plasmon excitations with angular  momenta larger than 1
substantially alter the  profile for multiphoton absorption
as compared to the single-photon case.

Our model is formulated in terms of a  charge density distribution
function $\rho(\bf r)$ for the cluster. This means that, in principle,
one can study the response for different charge density profiles
including deformed ones.
Our model is a semi-classical one, which neglects the granularity
of charge in the system. This is consistent with the principles
underlying the jellium picture. It is appropriate for metallic
clusters and, to a lesser extent, for fullerenes.

In the classical formulation of our model,
we have used Euler's equation for hydrodynamic
flow,  together with the equation of continuity.
We have demonstrated that the results following from our model
are consistent with direct estimates of the
matrix elements for the multiphoton absorption process.
The theoretical formalism we have developed is not confined in
its application to photons. It can also be used to describe
any kind of higher order plasmon excitation processes, for example
those which arise by multiple scattering of electrons within a cluster.

\section{Acknowledgments}

The authors acknowledge support from the
Royal Society of London, INTAS, the DAAD and
the Alexander von Humboldt Foundation.

\appendix

\section{
Matrix elements of plasmon resonance transitions
\label{coll_matr_el}
}

In this appendix, we evaluate the matrix elements
of plasmon resonance transitions in the plasmon resonance approximation
by the use of the sum rule.

For a stationary (i.e. explicitly independent of time) operator $\hat F$
of an observable physical quantity, characterising a system of particles
with the Hamiltonian $\hat H$, one can formulate the following sum rule
(see e.g. \cite{LL3})
\begin{equation}
\sum_n \omega_{n0} \left| \langle n| \hat F | 0 \rangle\right| ^2 =
\frac{1}{2}\langle 0 | \left[ \hat F, \left[\hat H, \hat F\right] \right] |0\rangle
\label{gen_sum_rule}
\end{equation}

Here  $\omega_{n0}= \varepsilon_n -\varepsilon_0$, the summation is
performed over all excited states of the system and
$\left[\hat H, \hat F \right]$
denotes the commutator of the operators
$\hat H$ and $\hat F$: $\left[\hat H, \hat F \right] = \hat H \hat F -\hat F \hat H $.

Applying the sum rule (\ref{gen_sum_rule}) to operator $\hat F$ defined as
\begin{equation}
\hat F= \sum_k F({\bf r}_k),
\label{F_operator}
\end{equation}
\noindent
where $F({\bf r}_k)$ is a function of the coordinates of the $k-th$ electron and
the summation in (\ref{F_operator}) is performed over all particles in
the system, one derives
\begin{equation}
\sum_n \omega_{n0} \left| \langle n| \hat F | 0 \rangle\right| ^2 =
\frac{\hbar^2}{2m}
\int d{\bf r} \left| {\bf \nabla} F({\bf r})\right|^2 \rho({\bf r})
\label{gen_sum_rule_1}
\end{equation}

Here $\rho({\bf r})$ is the ground state charge density distribution
in the system.  Applying the general rule (\ref{gen_sum_rule_1})
to the function
\begin{equation}
F({\bf r})=\sqrt{\frac{4\pi}{2l+1}} r^l Y_{lm}({\bf n})
\label{def_F}
\end{equation}
\noindent
and performing the integration in (\ref{gen_sum_rule_1}) with the density
distribution (\ref{rho_ini}), one derives
\begin{equation}
\sum_n \omega_{n0} \left| \langle n| \hat Q_{lm} | 0 \rangle\right| ^2 =
\frac{\hbar^2}{2} \omega_l^2 R^{2l+1}
\label{mult_mom_sum_rule}
\end{equation}

Here, we have used the following definition of the multipole moments operator
\begin{equation}
\hat Q_{lm}=\sqrt{\frac{4\pi}{2l+1}}
\sum_{k=1}^N e_k r_k^l Y_{lm}({\bf n_k}).
\label{Q_lm}
\end{equation}
\noindent
The plasmon  resonance frequencies in (\ref{mult_mom_sum_rule})
are  defined according to (\ref{omega_pl}).

Using (\ref{mult_mom_sum_rule}), one can easily evaluate
the matrix elements of plasmon resonance transitions in
the plasmon resonance approximation. Indeed,  assuming that
plasmon excitations dominate in the sum  over $n$
in (\ref{mult_mom_sum_rule}), one derives
\begin{equation}
Q_{lm}= \langle n|\hat Q_{lm}|0\rangle=
\sqrt{\frac{\hbar\omega_l R^{2l+1}}{2}}=
e R^{l-1} \sqrt{\frac{N\hbar}{2m\omega_l}}  \sqrt{\frac{3l}{2l+1}}
\label{Q_lm_fin}
\end{equation}

Equation (\ref{Q_lm_fin}) gives the matrix elements
of plasmon resonance transitions for an arbitraty large
angular momentum $l$.  The correctness of the result (\ref{Q_lm_fin})
can be independently varified by performing calculations of the
multipole dynamic polarizabilty of the cluster in the plasmon
resonance approximation. Indeed, using (\ref{Q_lm_fin}), one derives
\begin{equation}
\alpha_{l}(\omega)=  2\sum_n \frac{\omega_{n0}|Q_{l0}|^2}
{\omega_{n0}^2 -\omega^2 - i\omega \Gamma_l}
\approx
\frac{R^{2l+1} \omega_l^2}
{\omega_{l}^2 -(\hbar\omega)^2 - i\hbar^2 \omega \Gamma_l}
\end{equation}
\noindent
which is the known expression for the dynamic multipole
plasmon polarizability in the plasmon resonance approximation
(see e.g. \cite{Kreibig95,Gerchikov97a}).

For the dipole plasmon resonance transition, one derives
from (\ref{Q_lm_fin})
\begin{equation}
Q_{10}= d_{10} = e z_{10}= e \sqrt{\frac{N\hbar}{2m\omega_1}}
\label{Q_10}
\end{equation}
\noindent
which is consistent with the dipole sum rule (\ref{sum_rule}).

Equation (\ref{Q_lm_fin}) allows one to evaluate
the matrix elements of electronic transitions between
various plasmon resonance states. To demonstrate this,
let us rewrite (\ref{Q_lm_fin}) in the form
\begin{equation}
Q_{lm}= \sqrt{\frac{1}{2l+1}}  \int dr r^{l+2}
\rho_{l0}(r)
\label{Q_lm_rho}
\end{equation}

Here, we have introduced the radial transition density $\rho_{l0}(r)$
between the ground state and the excited plasmon resonance state
with angular momentum $l$
and used the relationship
$I_1(l_n,m_n|l,m|0,0)=\delta_{l_n,l} \delta_{m_n,m}/\sqrt{4\pi}$, when
calculating angular intergrals in (\ref{Q_lm_rho}).

The radial transition density $\rho_{l0}(r)$  is localised in the vicinity
of the cluster surface. Qualitatively, this is clear because $\rho_{l0}(r)$
describes the plasmon excitation. Quantitatively, this was proved
by {\it ab initio} computations of the transition densities in
the $Na_{40}$ and $Na_{92}$ clusters within the jellium model
in \cite{Gerchikov98}. Therefore, to a reasonable accuracy,
one can approximate $\rho_{l0}(r)$ by the delta function
\begin{equation}
\rho_{l0}(r)= \rho_{l0}\delta (r-R)
\label{rho_l0}
\end{equation}

Substituting (\ref{rho_l0}) into (\ref{Q_lm_rho}) and comparing
the result of the calculation with (\ref{Q_lm_fin}), one can
determine the value  $\rho_{l0}$ entering (\ref{rho_l0}).
The result of this calculation reads as
\begin{equation}
\rho_{l0}= \frac{e}{R^3}\sqrt{\frac{3l\hbar N}{2m\omega_l}}
\label{rho_l0_fin}
\end{equation}

Let us now evaluate the matrix element for the dipole
transition between  plasmon resonance modes, which
reads as
\begin{equation}
z_{l_2 l_1}=\sqrt{\frac{4\pi}{3}} \int d{\bf r}
Y_{1,0}({\bf n}) r \rho_{l_2 l_1}({\bf r})
\label{z12_def}
\end{equation}

Here, $\rho_{l_2 l_1}({\bf r})$  is the electron transition density between
the dipole and the quadrupole plasmon modes. This transition density
can be evaluated via the transition densities $\rho_{l_2, l_1}(r)$,
$\rho_{l_20}(r)$ and the ground state electron density
of the cluster $\rho_{0 0}=e N/V$ as follows
\begin{equation}
\rho_{l_2l_1}({\bf r})=\frac{\rho_{l_20}(r)\rho_{l_10}(r)}{\rho_{00}}
Y_{l_1,0}({\bf n}) Y_{l_2,0}({\bf n})
\label{rho_21}
\end{equation}

Substituting (\ref{rho_21}) into (\ref{z12_def}) and performing
simple transformations, one derives
\begin{equation}
z_{l_2 l_1} = e \sqrt{\frac{4\pi l_1 l_2}{3}}
\left(\frac{l_1(2l_2+1)}{l_2 (2l_1+1)} \right)^{1/4}
I_1(l_2,m_2|1,0|l_1,m_1)
\frac{2\pi \hbar}{m \omega_1 } \delta(0)
\label{z21_inter}
\end{equation}

This equation  has an uncertainty, which originates from the
fact that we have assumed zero width for the domain
in the vicinity of the cluster surface in which plasmon excitations
take place. By introducing a finite width $\Delta R$ for this domain
and using one of the standard representations of the $\delta$-function
\cite{LL3} to resolve the uncertainty,
$\delta(0) \approx 2/\pi \Delta R$,
one finaly derives
\begin{equation}
z_{l_2 l_1} = 8 e \sqrt{\frac{\pi l_1 l_2}{3}}
\left(\frac{l_1(2l_2+1)}{l_2 (2l_1+1)} \right)^{1/4}
I_1(l_2,m_2|1,0|l_1,m_1)
\frac{\hbar}{m \omega_1 \Delta R}
\label{z21_final}
\end{equation}

The explicit expression for the angular intergral
$I_1(l_2,m_2|1,0|l_1,m_1)$ is given in \ref{integrals}.
In the case of the transition between  the dipole
and quadrupole plasmon resonance states this integral
is equal to $I_1(2,0|1,0|1,0) = - 1/\sqrt{5\pi}$.
Substituting this value
into (\ref{z21_final}) and performing
simple algebraic transformations, one arrives at the  expression for
the  matrix element describing the transition between the dipole and
the quadrupole plasmon resonance modes.
\begin{equation}
z_{12}= - e\frac{8}{3} \left(\frac{6}{5}\right)^{1/4}
\frac{\hbar}{m\omega_1\Delta R}
\label{z12_fin}
\end{equation}

\section{
Integrals $I_1(l,m|l_1,m_1,|l_2,m_2)$ and $I_2(l,m|l_1,m_1,|l_2,m_2)$
}
\label{integrals}

The angular integral,
\begin{equation}
I_1(l,m|l_1,m_1|l_2,m_2)=
\int d\Omega_{{\bf n}_r}
Y_{l,m}^{*}({\bf n}_r) Y_{l_1,m_1}({\bf n}_r) Y_{l_2,m_2}({\bf n}_r),
\label{A1}
\end{equation}
is well known and can be found in many textbooks (see e.g. \cite{VMH}).
It is equal to

\begin{eqnarray}
I_1(l,m|l_1,m_1|l_2,m_2)&=&
(-1)^m i^{l_1+l_2-l} \sqrt{\frac{(2l+1)(2l_1+1)(2l_2+1)}{4\pi}}\times
\nonumber
\\
&\times&
\threejot{  l}{ l_1}{l_2}
         {-m }{ m_1}{ m_2}
\threejot{ l}{l_1}{l_2}
         {0 }{  0}{  0}
\label{A2}
\end{eqnarray}

Here, the integral is expressed via 3j-symbols (for definition
see  e.g. \cite{VMH}).

The angular integral,

\begin{eqnarray}
I_2(l,m|l_1,m_1|l_2,m_2)&=&
\sqrt{l_1(l_1+1)l_2(l_2+1)}
\int d\Omega_{{\bf n}_r} Y_{l,m}^{*}({\bf n}_r)\times
\nonumber
\\
&& \times  {\bf Y}^{(1)}_{l_1,m_1}({\bf n}_r)
{\bf Y}^{(1)}_{l_2,m_2}({\bf n}_r),
\label{A3}
\end{eqnarray}
can be expressed via
the sum of products of 3j-symbols and 6j-symbols (for definition
see  e.g. \cite{VMH}) after performing the following
transformations.

Using the standard relationships for spherical vector harmonics,
written in \cite{VMH} on page 210, one derives
\begin{eqnarray}
I_2(l,m|l_1,m_1|l_2,m_2)=
\sqrt{l_1(l_1+1)l_2(l_2+1)}
\int d\Omega_{{\bf n}_r} Y_{l,m}^{*}({\bf n}_r)\times
\nonumber
\\
\times
\left(\sqrt{\frac{l_1+1}{2l_1+1}}{\bf Y}^{l_1-1}_{l_1,m_1}({\bf n}_r)
+ \sqrt{\frac{l_1}{2l_1+1}}{\bf Y}^{l_1+1}_{l_1,m_1}({\bf n}_r)\right)
\times
\nonumber
\\
\times
\left(\sqrt{\frac{l_2+1}{2l_2+1}}{\bf Y}^{l_2-1}_{l_2,m_2}({\bf n}_r)
+ \sqrt{\frac{l_2}{2l_2+1}}{\bf Y}^{l_2+1}_{l_2,m_1}({\bf n}_r)\right)
\label{A4}
\end{eqnarray}
The integrations arising in (\ref{A4}) can be performed and expressed
via the sum of products of 3j-symbols and 6j-symbols, using
the standard formulae given in \cite{VMH} on pages 222 and 236.
The result of the calculations of these integrals reads as

\begin{eqnarray}
I_2(l,m|l_1,m_1|l_2,m_2)=
\nonumber
\\
=(-1)^{l_1+l_2+l+m+1} \sqrt{l_1(l_1+1)l_2(l_2+1) (2l+1)}
\threejot{l_1}{l_2}{l}
       {m_1}{m_2}{-m}
\times
\nonumber
\\
\times
\Biggl\{
\sqrt{\frac{(l_1+1)(l_2+1)(2l_1-1)(2l_2-1)}{4\pi}}
\threejot{l_1-1}{l_2-1}{l}
         {0}{ 0}{0}
\sixjot{l_1-1}{l_2-1}{l}
       {l_2}{l_1}{1}+
\nonumber
\\
+
\sqrt{\frac{l_1(l_2+1)(2l_1+3)(2l_2-1)}{4\pi}}
\threejot{l_1+1}{l_2-1}{l}
         {0}{ 0}{0}
\sixjot{l_1+1}{l_2-1}{l}
       {l_2}{l_1}{1}+
\nonumber
\\
+\sqrt{\frac{(l_1+1)l_2(2l_1-1)(2l_2+3)}{4\pi}}
\threejot{l_1-1}{l_2+1}{l}
         {0}{ 0}{0}
\sixjot{l_1-1}{l_2+1}{l}
       {l_1}{l_2}{1}+
\nonumber
\\
+\sqrt{\frac{l_1 l_2(2l_1+3)(2l_2+3)}{4\pi}}
\threejot{l_1+1}{l_2+1}{l}
         {0}{ 0}{0}
\sixjot{l_1+1}{l_2+1}{l}
       {l_2}{l_1}{1}
\Biggr\}
\label{A5}
\end{eqnarray}


For the particular cases of interest, one derives from (\ref{A5})
\begin{eqnarray}
I_2(1,0|0,0|1,0)&=& 0
\nonumber
\\
I_2(0,0|1,0|1,0)&=& \frac{1}{\sqrt{\pi}}
\nonumber
\\
I_2(2,0|1,0|1,0)&=& \frac{1}{\sqrt{5\pi}}
\nonumber
\\
I_2(1,0|2,0|1,0)&=& \frac{6}{10\sqrt{\pi}}\left(1+ \frac{16}{3\sqrt{5}}  \right)
\nonumber
\\
I_2(3,0|2,0|1,0)&=& - \frac{18}{5\sqrt{105\pi}}\left(1+ \frac{\sqrt{5}}{2} \right)
\label{A6}
\end{eqnarray}




\end{document}